\documentclass[nohyper]{JHEP3}

\usepackage{graphicx}
\usepackage{epsfig,cite}

\newcommand{\be}{\begin{equation}}
\newcommand{\ee}{\end{equation}}
\newcommand{\bea}{\begin{eqnarray}}
\newcommand{\eea}{\end{eqnarray}}
\newcommand{\bi}{\begin{itemize}}
\newcommand{\ei}{\end{itemize}}
\newcommand{\ben}{\begin{enumerate}}
\newcommand{\een}{\end{enumerate}}
\newcommand{\la}{\left\langle}
\newcommand{\ra}{\right\rangle}
\newcommand{\lc}{\left[}
\newcommand{\rc}{\right]}
\newcommand{\lp}{\left(}
\newcommand{\rp}{\right)}

\newcommand{\mrexp}{\mathrm{exp}}
\newcommand{\dat}{\mathrm{dat}}
\newcommand{\stopp}{\mathrm{stop}}
\newcommand{\gen}{\mathrm{gen}}
\newcommand{\art}{\mathrm{art}} 
\newcommand{\rep}{\mathrm{rep}}
\newcommand{\net}{\mathrm{net}}

\newcommand{\stat}{\mathrm{stat}}

\newcommand{\tot}{\mathrm{tot}}

\newcommand{\tth}{\mathrm{th}}
\newcommand{\ew}{\mathrm{ew}}
\newcommand{\pert}{\mathrm{pert}}
\newcommand{\draft}[1]{}

\title{Neural network parametrization
of the lepton energy spectrum in
semileptonic B meson decays}

\author{Joan Rojo\\
        Departament d'Estructura i Constituents de 
        la Mat\`eria,\\
        Universitat de Barcelona, Diagonal 647, E-08028 Barcelona, Spain. \\
        E-mail: \email{joanrojo@ecm.ub.es}}

\abstract{We construct a parametrization of the
lepton energy spectrum in inclusive semileptonic decays of
B mesons, based on the available experimental information: moments of the
spectrum with cuts, their errors
and their correlations, together with kinematical
constraints. The result is obtained in the form
of a Monte Carlo sample of neural networks trained on 
replicas of the experimental data, which represents the probability
density
in the space of lepton energy spectra. This parametrization
is then used to extract the b quark mass $m_b^{1S}$ in a way
that theoretical uncertainties are minimized, for which the value
  $m_b^{1S}=4.84\pm 0.14^{\exp}\pm 0.05^{\tth}$ GeV is obtained.}


\preprint{hep-ph/0601229}

\begin{document}


\section{Introduction and motivation}

In the last decade the field of B meson physics has been the object
of a wealth of studies (see Ref. \cite{superB} and references therein),
motivated by the high precision measurements
 from the B factories, Belle and Babar. In particular,
the inclusive semileptonic decays $B\to X l\nu$, where $X$ stands for
a hadronic system, have received a lot of
attention, both in the theoretical\cite{gambinomom,trott} 
and in the experimental sides (see \cite{gambinorev} for
an up-to-date summary of the present situation), due
to its paramount importance for the determination the CKM matrix elements,
 and also since they provide  important
information on the underlying strong interaction dynamics.

It is well known that differential distributions in inclusive 
semileptonic decays
of heavy mesons can be computed by means of the Operator Product
Expansion \cite{ope1,ope2}. The resulting distributions are singular and can 
only be compared with the experimentally measured distributions
after smearing over a sufficiently large interval.

In principle one can measure not only the branching ratios
of these modes but also the full differential spectra on certain kinematical
variables, like the lepton energy or the hadronic invariant mass. However,
practical considerations force that the 
observables that are measured are convolutions of these
spectra with suitable weight functions and given kinematical cuts.
The most common case is when this observables are moments of the
spectra. 
On the other side, as has been mentioned before,
there is no pointwise theoretical prediction
for these spectra, since the output of the theoretical
computation is not a normal function but rather
a distribution, which  is a general
feature of partonic cross sections, and, therefore, only integrals over 
a sufficiently large energy range can be 
reliably computed in perturbation theory. 
Therefore, one has to smear the theoretical prediction for
the spectrum to compare with the experimental measurements.
 
From all the above reasons, it is clear that 
it would be interesting to obtain from  experimental data
the full spectrum with uncertainties, to allow a more
general comparison 
 with theoretical predictions. 
Such parametrization of the spectra would, for example, allow
a comparison of general convolutions of the spectra with arbitrary
kinematical cuts with theoretical computations, even if these
convolutions have not been measured experimentally.
 Another application could
be to study possible violations of quark-hadron duality
in these lepton spectra, or to estimate the size of 
higher order corrections, both perturbative and nonperturbative.

Our purpose in this work is to allow for a
 more general comparison of the
theoretical predictions with the experimental data. With this aim 
a parametrization of the lepton energy spectrum from available
experimental information on its moments is constructed, supplemented by 
constraints from the kinematics of the process. 
Traditional strategies, like fits with functional forms, suffer
from the well known problem of parametrization bias  and, 
moreover, do not allow a determination of the  uncertainties
associated 
to the parametrization, so new suitable strategies must be
developed to address this problem in a statistically
meaningful way.

Recently, a novel approach to the problem of the
parametrization of experimental data in an unbiased way with a faithful
estimation of the uncertainties was proposed, based on
the combination of Monte Carlo techniques and neural networks
as basic interpolating tools, which determines the probability 
density of the parametrization.
This technique has been successfully applied to
the parametrization of deep-inelastic structure functions
\cite{f2ns,f2nnp}, spectral functions from hadronic tau decays
\cite{tau} and parton distribution 
functions\footnote{Many ideas that appear in the present work will be
developed in a more detailed way in a
forthcoming publication \cite{nnqns}.} \cite{pdf,heralhc}.

This success motivated us to implement this technique
in the context of B physics. Therefore, in this work we
construct an unbiased parametrization of the lepton energy spectrum is
semileptonic B meson decays with a faithful estimation
of the uncertainties.
Since in Refs. \cite{f2ns,f2nnp,tau,pdf} 
the technique that will be used in this
work is  discussed in detail, here only those
aspects which are special to the present application will be
emphasized. Fig. \ref{graph} shows a diagram that summarizes our 
parametrization strategy.

\FIGURE[ht]
{\epsfig{width=1.0\textwidth,figure=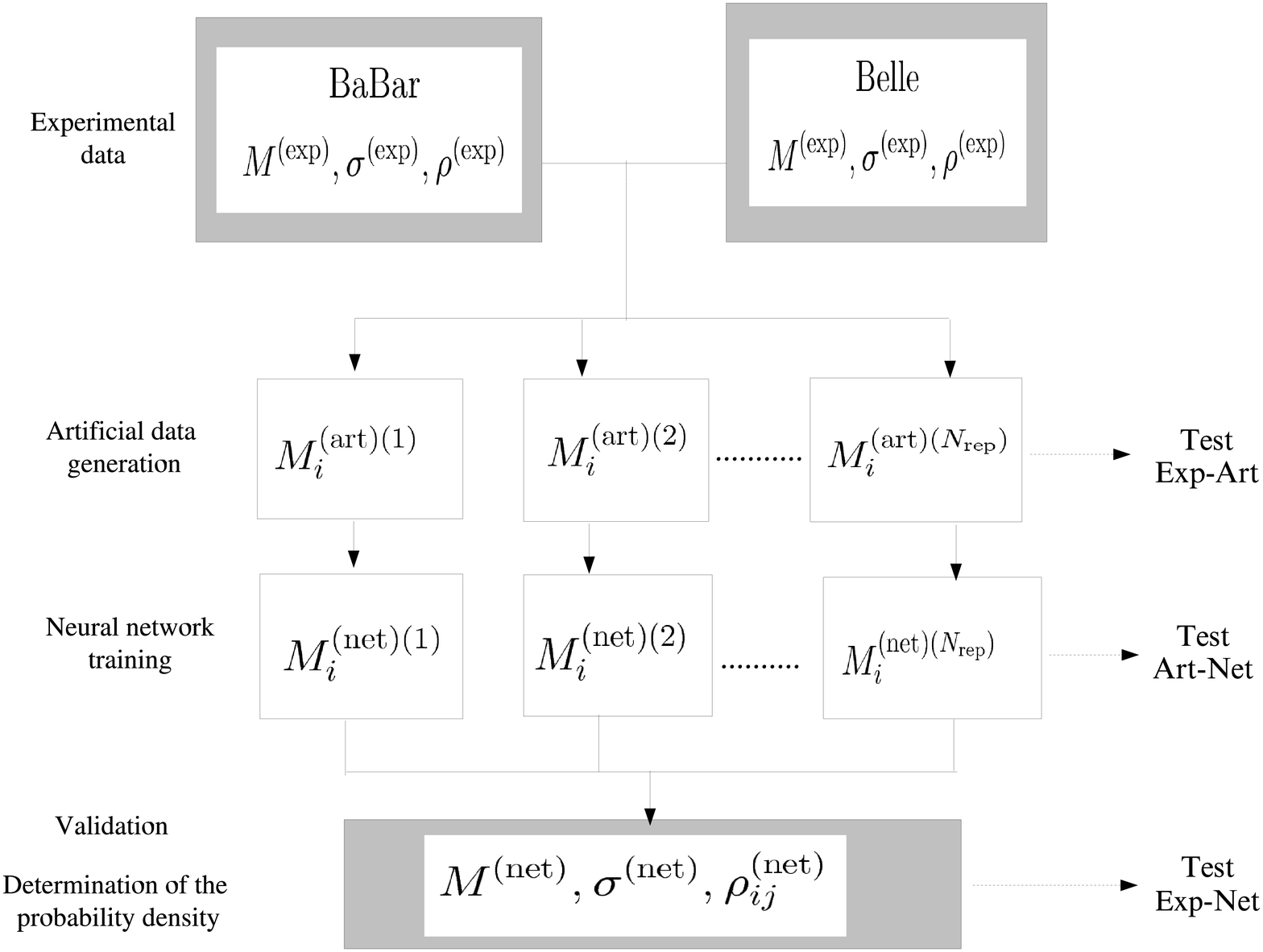} 
        \caption{Diagram that represents schematically
our strategy to parametrize the lepton energy spectrum with a 
Monte Carlo sample of neural networks.}
        \label{graph}}

As a byproduct of our analysis an extraction of the
heavy quark nonperturbative parameters $\bar{\Lambda}_{1S} $ and $\lambda_1$
will be performed using a technique that ensures
that the associated theoretical uncertainties are minimized \cite{bauertrott}.
It will be shown that in this case 
the dominant uncertainties in the determination
of these parameters turn out to be
the experimental uncertainties, that is, those associated to
the uncertainties in the parametrization of the spectrum.

Summarizing, there are three main motivations to construct a neural
network parametrization of the
lepton energy spectrum in B meson decays. The first one
is to generalize the approach of Refs. \cite{f2ns,f2nnp,tau,nnqns}
to the problem of the construction of a unbiased determination of
physical quantities with faithful estimation of their uncertainties
from experimental data in the case for which the only available
information on this quantity comes through truncated moments,
as is the case for the lepton energy spectrum. Second, 
to show how this parametrization allows a more general comparison
of theoretical predictions with data, since from our
parametrization one can extract for example moments that
have not been measured, like non-integer moments, higher
order moments or moments with large cut in the lepton
energy $E_0$, and use them for several purposes. 
In this work we examine two of such applications: the
comparison of the theoretical accuracy with which higher
order moments or moments with large $E_0$ are computed with
respect to that of experimental measurements
(Section 6.2),
or novel methods to determine non perturbative parameters
like $\overline{m}_b\lp \overline{m}_b\rp$ 
from non-integer moments (Section 7). 
Finally, the set of techniques
described in the present work allow for a straightforward generalization
to other relevant problems in B meson physics, like
the determination of the B meson shape function from experimental data
$S(\omega)$ \cite{shapefun}.

The outline of this paper is as follows: in Section 2 we summarize
the theoretical aspects of semileptonic B meson decays, and in Section
3 the experimental data that will be used. In Section 4 we describe the
generation of Monte Carlo replicas of the experimental data, and in
Section 5 the process of neural network training. In Section 6
we present the results that are obtained for the lepton energy
spectrum and in Section 7 the 
determination of the nonperturbative parameters
 $\bar{\Lambda}_{1S} $ and $\lambda_1$.
Finally, in Section 8 we conclude and  briefly sketch
possible new applications of our strategy to other problems
in the context of B meson physics. Two
appendices summarize the most technical details
of the neural network parametrization. 

\section{Theory overview}

\FIGURE[ht]{\epsfig{width=0.80\textwidth,figure=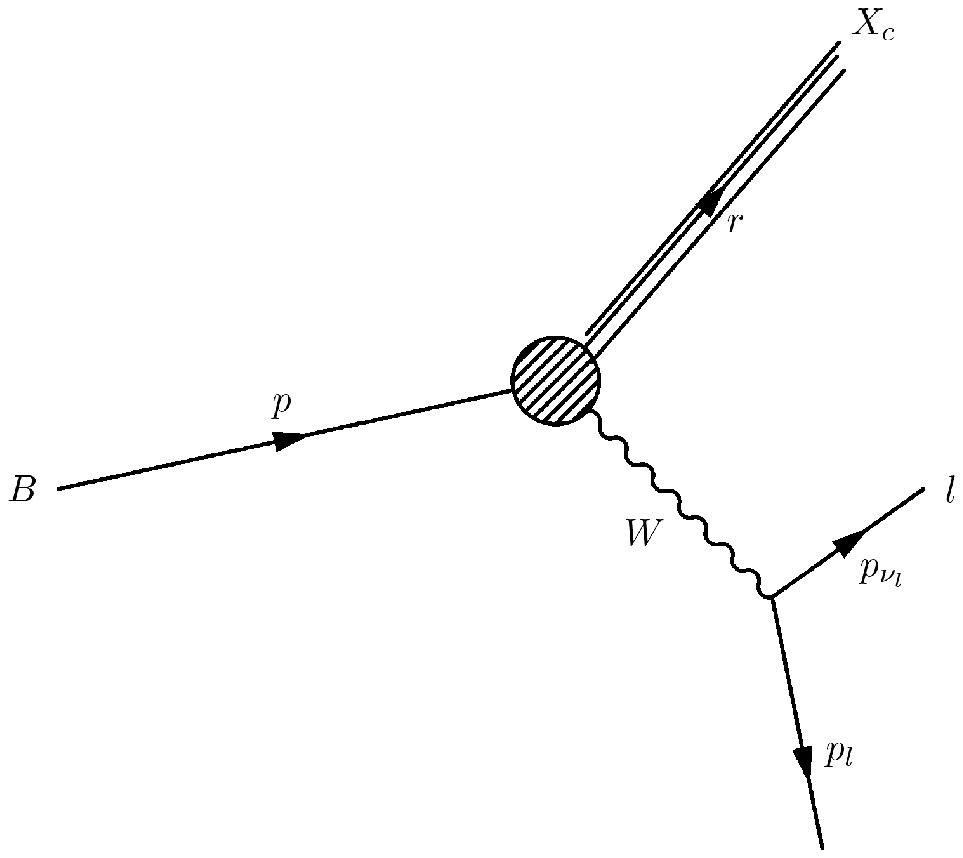}
\caption{Semileptonic decay of a B meson into a charmed
final state}
\label{bdecay}
}

In this work
the inclusive semileptonic decays of $B$ mesons with charmed
final states will be considered.  
Therefore, the process that will be analyzed  is
$B\to X_c l\nu$, which is represented in Fig. \ref{bdecay}.
The differential decay rate for this
process,
\be
B(p) \to l(p_l)+\bar{\nu}(p_{\bar{\nu}})+X_c(r) \ ,
\ee
depends on three different kinematical variables $q^2,r$ and $E_l$,
where $q=p_l+p_{\nu}$ is the total four 
momentum of the leptonic system, $r=p-q$ is the
four-momentum of the charmed hadronic final state, 
with invariant mass $r^2=M_X^2$, and $E_l$ is the lepton
 energy in the rest frame of the
 B meson. This triple differential distribution
 can be decomposed, taking into account
the kinematics of the process and the symmetries
of the theory,
 in terms of three structure functions $W_i$, 
\bea
\label{triple}
\frac{d^3\Gamma}{dq^2 dr dE_l}(q^2,r,E_l)=\frac{G_F^2|V_{cb}|^2}{16\pi^4}
\Bigg[ \hat{q}^2W_1(\hat{q}^2,\hat{u})-\qquad \qquad \nonumber \\
\lp 2v\cdot\hat{p}_l-2v\cdot\hat{p}_lv \cdot\hat{q}+\frac{\hat{q}^2}{2}\rp 
W_2(\hat{q}^2,\hat{u})+\hat{q}^2
\lp 2v\cdot \hat{p}_l-v \cdot \hat{q}\rp W_3(\hat{q}^2,\hat{u})\Bigg] \ ,
\eea
where $u^2=r^2-m_c^2$, $v=p/m_b $, and
 the quantities with a hat are dimensionless quantities
normalized to $m_b$. All the structure functions $W_i(\hat{q}^2,\hat{u})$ 
have
both a perturbative expansion in powers of $\alpha_s$, and
a nonperturbative expansion in powers of $1/m_b$, which
can be computed in the framework of the heavy quark expansion. 
For example,
the  complete set of $\mathcal{O}(\alpha_s)$ 
corrections
for all the differential distributions that can be constructed
from Eq. \ref{triple} with arbitrary kinematical cuts
 have recently become
available  \cite{trottmom,agru}.

The most general 
observables that are accessible from the experiments, as it will
be discussed below, 
are convolutions of  differential distributions
with suitable weight functions over a large enough
range of energy, with kinematical cuts. A particular case of these 
observables
are the  moments of differential
 decay distributions. In this work the focus will be on leptonic 
moments, defined as
\be
\label{ln}
L_n(E_0,\mu)\equiv \int_{E_0}^{E_{\max}} dE_l \lp E_l -\mu \rp^n 
\int dq^2 dr \frac{d^3\Gamma}{dq^2 dr dE_l}(q^2,r,E_l) \ ,
\ee
where $E_0$ is a lower cut on the lepton energy, 
and $E_{\max}$ is the maximum energy
allowed from the kinematics of the process that the
lepton can have,
\be
\label{emax}
E_{\max}=\frac{m_B^2-m_D^2}{2m_B} \ ,
\ee
where $m_B$ is the average of the mass of the neutral and charged
B mesons, and similarly for $m_D$. The lower cut in the lepton
energy in Eq. \ref{ln} is imposed by  experimental
requirements,
as will be discussed in the next section.
The quantity that is going to be
parametrized with a Monte Carlo sample of neural networks, the
lepton energy distribution, is defined as
 \be  
\frac{d\Gamma}{dE_l}(E_l)\equiv
\int dq^2 dr \frac{d^3\Gamma}{dq^2 dr dE_l}(q^2,r,E_l) \ ,
\label{specdef}
\ee
that is related to the observable leptonic moments via
\be
\label{leptonmom}
L_n(E_0,\mu)=\int_{E_0}^{E_{\max}} dE_l \lp E_l -\mu \rp^n 
\frac{d\Gamma}{dE_l}(E_l) \ .
 \ee

The lepton energy spectrum, Eq. \ref{specdef}, in $B\to X_c l\nu$
decays, as has been discussed before,
can  be expanded in a power series both
in $\alpha_s$ and in $1/m_b$. The leading order spectrum in both
expansions  
is given by \cite{ope2}
\be
\label{dgde}
\frac{d\Gamma}{dy}\lp B\to X_cl\nu\rp=\Gamma_0
2y^2\lc (1-f)^2(1+2f)(2-y)+(1-f)^3(1-y)\rc\theta(1-y-\rho) \ ,
\ee
\be
f=\frac{\rho}{1-y},\qquad \rho=\frac{m_c^2}{m_b^2},\qquad
y=\frac{2E_l}{m_b} \ ,
\ee 
where 
\be
 \Gamma_0=\frac{G_F^2m_b^5}{192\pi^3} 
\ee 
is the total parton model decay rate in the approximation of
a massless final state quark.
The kinematic support of the spectrum at this
leading order partonic level is 
\be
E_l\in \lc 0,\frac{m_b^2-m_c^2}{2m_b}\rc \ , 
\ee
where the upper limit is modified by nonperturbative
(hadronic) corrections.
The leading perturbative $\mathcal{O}(\alpha_s)$ 
corrections to this spectrum have been known for some
time \cite{kuhn}, and there are estimations of the size of higher
order terms though the BLM expansion \cite{blm}.
The leading nonperturbative $\mathcal{O}(1/m_b^2)$ 
corrections to the lepton
energy spectrum where computed in Refs. \cite{ope2,manohar}
and the  $\mathcal{O}(1/m_b^3)$ corrections in Ref. \cite{gremm}. 

The total decay rate, obtained by integration
of Eq. \ref{specdef}, admits the following  heavy quark 
expansion \cite{ope2,bigitotal}: 
\bea
\label{gammamb3}
\Gamma\lp B\to X_c l\nu\rp=\Gamma_0
|V_{cb}|^2\lp1+A_{\ew}\rp A_{\pert} \cdot\quad \nonumber \\
\Bigg[ z_0(\rho)\lp 1+\frac{\lambda_1}{2m_b^2}\rp 
+g(\rho)
\frac{\lambda_2}{2m_b^2}+\mathcal{O}\lp \frac{1}{m_b^3}\rp \Bigg] \ ,
\eea
which depends up to $\mathcal{O}\lp1/m_b^2\rp$
on the the nonperturbative 
parameters $\lambda_1$ and $\lambda_2$, and 
where  the phase space factors are given by
\be
\label{zo}
z_0(\rho)\equiv 1-8\rho+8\rho^3-\rho^4-12\rho^2\log \rho, \quad 
\rho=\frac{m_c^2}{m_b^2} \ ,
\ee
\be
g(\rho)=-9+24\rho-72\rho^2+72\rho^3-15\rho^4-36\rho^2\ln \rho \ ,
\ee
and where $A_{\ew}$ stands for the
electroweak corrections and $A_{\pert}$ for the QCD perturbative
corrections.
Similar heavy quark expansions are available
 for the lepton energy moments (see
 Ref. \cite{gambinomom} and references therein).

\section{Experimental data}
The experimental data that will be used in the
present analysis consists on
moments with kinematical cuts
of the lepton energy  distribution in  semileptonic
B meson decays to charmed final states $B\to X_c l\nu$.
These moments have recently been measured with great accuracy at the 
B factories, Babar \cite{Aubert:2004td} and Belle \cite{elmombelle}, as well as
by Cleo \cite{cleoleptonmom}. Therefore, in
 the present analysis the latest data from these
three experiments will be used. 
Data from CDF \cite{cdfhadmom} is not
incorporated since it is
restricted to hadronic moments.

As it has been mentioned before,
the main experimental difficulty
for the measurement of  the
lepton energy spectrum for low values
of the lepton energy is the fact that for low lepton
energies the
background from other decay modes dominates, and it
is challenging to disentangle the desired decay mode.
Therefore, kinematical cuts have to be imposed
 that remove
the low $E_l$ region of the spectrum.
Another relevant consideration is that
the reference frame
change, from the laboratory frame to the
B meson rest frame and several experimental 
corrections, like for
example electroweak final state radiation,
are easier to perform in terms
of moments of the distribution.
Therefore the final published measurements are moments of the lepton
energy spectrum, Eq. \ref{leptonmom}, 
with different cuts in the lepton energy, rather than the
full spectrum itself.

Now the data that will be used  for the present 
parametrization of the lepton energy  spectrum
will be described.
The Babar Collaboration \cite{Aubert:2004td}
provides  the partial branching ratios,
\be
\label{momexp1}
M_0(E_0)=\tau_BL_0(E_0,0)=\tau_B\int_{E_0}^{E_{\max}}
\frac{d\Gamma}{dE_l}(E_l)~dE_l \ ,
\ee
where $\tau_B$ is the average B meson lifetime \cite{hfag},
the first moment,
\be
M_1(E_0)=\frac{L_1(E_0,0)}{L_0(E_0,0)} \ ,
\label{babmom}
\ee
and the central moments,
\be
\label{momexp2}
M_n(E_0)=\frac{L_n(E_0,M_1(E_0))}{L_0(E_0,0)}, \qquad n=2,3 \ ,
\ee
for five different values of $E_0$ from 0.6 to 1.5 GeV,
which makes a total of 20 data points.
The rationale for extracting Eq. \ref{momexp2} rather than
for example
\be
\widetilde{M}_n(E_0)=\frac{L_n(E_0,0)}{L_0(E_0,0)}, \qquad n=2,3 \ ,
\ee
is that in the former case correlations between different
moments are smaller and therefore more independent information
can be extracted from the measurements.

The Belle Collaboration \cite{elmombelle} provides the
same moments, $M_n(E_0)$  for $n=1$, $n=2$ and $n=3$\footnote{
For example, they define $M_1=\la E_l\ra$, which if one
takes into account that the corresponding normalized
probability density is given by
\be
\mathcal{P}(E_l)=\lp\frac{1}{\int_{E_0}^{E_{\max}}\frac{d\Gamma}{dE_l}
dE_l}\rp\frac{d\Gamma}{dE_l}(E_l),\qquad E_0 \le E_l \le E_{\max}
\ee
one ends up with Eq. \ref{babmom}, and
similarly for the remaining moments.}.
The difference with the Babar data is that the partial branching
ratio Eq. \ref{momexp1} is not measured, and that the
Belle data cover a somewhat larger lepton energy range, since the
lowest value of $E_0$ of their data set is $E_0=0.4$ GeV.
These moments, for six  different values of $E_0$ from 0.4 to 1.5 GeV,
make up a total of 18 data points. 
Finally the Cleo Collaboration \cite{cleoleptonmom} provides 
the moments
$M_n(E_0)$ for
$n=1,2$, for energies between 0.6 to 1.5 GeV, for a total of
20 data points (10 data points for $n=1$ and 10 data points 
for $n=2$). The average 
correlations
for this experiment are larger since measurements of the same
moment at different energies $E_0$ are highly correlated.

The three collaborations provide also the total 
and statistical errors, as well as the
correlation between different measurements. These features
are summarized in Table \ref{datafeat1}. 
 Note that for all experiments
correlations are rather large, so it is compulsory to incorporate them
in a consistent way in the statistical analysis of the data.
However, one has
to be careful with the treatment of the  
experimental correlations, for  reasons to be described on the
next section.

Note that the results of this work, summarized in
section \ref{results}, consist on a parametrization
of experimental data without the need for 
any theoretical input. The uncertainties associated to
the parametrization of the lepton spectrum will
therefore be reduced if experimental measurements of
lepton spectrum moments are measured with larger
accuracy in the future.

\subsection{Treatment of experimental correlations}

\label{treatcor}

As has been already noticed, for example see the
global analysis of B decays of Ref. \cite{bauerglobalfit},
it can be checked that the experimental correlation
matrices, $\rho_{ij}^{(\mrexp)}$, 
as presented with the published data of the
three experiments \cite{elmombelle,Aubert:2004td,cleoleptonmom},
are not positive definite. The source of this problem can be traced to the
fact that
off-diagonal elements of correlation matrix
are large, as expected since moments with similar energy cuts
contain almost the same amount of information and are therefore
highly correlated. Then one can check that 
some eigenvalues are negative and small, 
and 
this points to the fact that the source of the problem is
an insufficient accuracy in the computation of the
elements of the correlation matrix.

However, whatever is the
original source of the problem, the fact that the
experimental correlation matrices
are not positive definite has an important consequence:  the
 technique introduced in
\cite{f2ns} for the generation of a sample of replicas of the
experimental data in a way that correlations
are incorporated 
 relies on the existence of a positive definite
correlation matrix, and therefore if this
is not the case our
technique cannot be applied. 

A method to overcome these difficulties while keeping the maximum
amount on information on experimental correlations as
possible consists on removing those data points for which
the experimental correlations are larger 
than a maximum correlation, $\rho_{ij}^{(\exp)}\ge \rho^{\max}$.
The value of $\rho^{\max}$ is determined separately for each
experiment as the maximum value for which the resulting
correlation matrix is positive definite. In Table
\ref{datafeat2}  the values of $\rho^{\max}$ for each
experiment are shown, 
together with the features of the remaining experimental data
after those data points with too large correlations have
been removed. In the case of the Belle measurements, 
another problem
with the correlation matrix is that the contribution
to the correlation coefficients from systematic uncertainties
has at  present not been incorporated. 

\TABLE[ht]{
\begin{tabular}{|c|c|cc|cc|c|}  
\hline
Experiment & $N_{\dat}$ &  $n$ &  $E_0$ (GeV) & 
$\la \sigma_{\stat}\ra$ &    $\la \sigma_{\tot}\ra$ &
$\la | \rho |\ra$ \\
\hline
Babar \cite{Aubert:2004td} & 20 & 0~-~3 & 0.6~-~1.5& 6.0\% & 8.0\% & 0.50   \\
Belle \cite{elmombelle} & 18 & 1~-~3 & 0.4~-~1.5& 15.0\% & 16.0\% & 0.34  \\
Cleo\cite{cleoleptonmom} & 20 & 1~-~2 & 0.6~-~1.5& 1.0\% & 1.3\% & 0.65  \\
\hline
\end{tabular}
\caption{\small Features of experimental data on lepton
moments $M_n(E_0)$.  Note that averages over 
experimental errors are given as percentages.
\label{datafeat1}}
}

\TABLE[ht]{
\begin{tabular}{|c|c|cc|cc|cc|}  
\hline
Experiment & $N_{\dat}$ &  $n$ &  $E_0$ (GeV) & 
$\la \sigma_{\stat}\ra$ &    $\la \sigma_{\tot}\ra$ &
$\rho^{\max}$ & $\la |\rho|\ra$ \\
\hline
Babar \cite{Aubert:2004td} & 16 & 0~-~3 & 0.6~-~1.5& 4.0\% & 5.0\% & 
0.97 & 0.49   \\
Belle \cite{elmombelle} & 15 & 1~-~3 & 0.4~-~1.5& 18.0\% & 19.0\% & 
0.88 & 0.31  \\
Cleo \cite{cleoleptonmom} & 10 & 1~-~2 & 0.6~-~1.5& 
0.5\% & 1.0\% &  0.95 &0.69  \\
\hline
\end{tabular}
\caption{\small Features of experimental data that is
included in the fit, after data points
with too large correlations have been removed. Note 
that averages over experimental
error are given as percentages. } 
\label{datafeat2}
}

\section{Replica generation}
As has been mentioned in the Introduction, in this work 
 the strategy 
of Ref. \cite{f2ns} is followed 
to parametrize the lepton energy spectrum from
the experimental information on its moments. The first step of this 
technique is to generate an ensemble of 
Monte Carlo replicas of the original
experimental data, which consists in the measured moments, which 
will be denoted by
\be
M_i^{(\mrexp)}, \quad  i=1,\ldots,N_{\dat} \ ,
\ee
where $M_i^{(\mrexp)}$ stands for any of Eqns. 
\ref{momexp1}-\ref{momexp2}, and $N_{\dat}$ is the 
total number of experimental data points,
together with the total error and the correlation matrix.

To generate replicas one proceeds as follows: 
the k$-$th artificial replica of the
experimental data $M^{(\art)(k)}$ is constructed as
\be
M_j^{(\art)(k)}=M_j^{(\mrexp)}+s^{(k)}_j\sigma^{(\mrexp)}_{j,\tot}  \ ,
\quad
j=1,\ldots,N_{\dat}, \quad, k=1,\ldots,N_{\rep}\ ,
\ee
where $s^{(k)}_j$ are gaussian random numbers with same correlation
matrix 
as the experimental correlation matrix $\rho_{ij}$,
 $\sigma^{(\mrexp)}_{j,\tot}$ is the total error of the
j-th data point and $N_{\rep}$ is the number of generated
replicas of the experimental data.
This way the ensemble of replicas is not only able to reproduce
the central values and the errors 
but also the correlations of the experimental data. 

As explained in Ref. \cite{f2ns}, the size of the replica sample
is fixed by the condition that the averages over replicas reproduce the
experimental values for central values, errors and correlations.
The different statistical estimators  are defined in
Appendix \ref{estimators}. In Table \ref{gendata} 
the relevant statistical estimators for the
replica generation are summarized. One can observe that
to reach the desired accuracy of a few percent
and to have scatter correlations $r\ge 0.99$
for central values, errors and correlations,
 a sample of 1000 replicas is 
needed.

\TABLE[ht]{
\begin{tabular}{|c|ccc|}  
\hline
 $N_{\rep}$ & 10 & 100 & 1000\\
\hline
$\la PE\lc\la M \ra_{\rep}\rc\ra$ & 2.47\%
& 0.40\% & 0.24\%   \\
\hline
$\la PE\lc \sigma^{(\art)}\rc\ra_{\dat}$ & 32.4\% 
& 13.8\% & 3.4\% \\
$\la \sigma^{(\art)}\ra_{\dat}$ &  0.00265 & 0.00277 & 0.00268\\
$r\lc \sigma^{(\art)}\rc$ &  0.95 & 0.99 & 0.99 \\
\hline
$\la PE \lc \rho^{(\art)} \rc \ra_{\dat}$ & 60.1\%
& 19.6\% & 6.7\%   \\
$\la \rho^{(\art)}\ra_{\dat}$ & 0.132 & 0.138 & 0.155 \\
$r\lc \rho^{(\art)}\rc$ &  0.75 & 0.96 & 0.99  \\
\hline
$\la \mathrm{cov}^{(\art)}\ra_{\dat}$ & $1.1~10^{-6}$ 
& $1.4~10^{-6}$  & $1.3~10^{-6}$ \\
$r\lc \mathrm{cov}^{(\art)}\rc$ & 0.86 & 0.98 & 0.99 \\
\hline
\end{tabular}
\caption{\small Comparison between experimental and 
Monte Carlo data.\hfill\break
The experimental data  have
$\la \sigma^{(\mrexp)}\ra_{\dat}=0.00267$ ,
 $\la \rho^{(\mrexp)}\ra_{\dat}=
0.166$ and $\la \mathrm{cov}^{(\mrexp)}\ra_{\dat}=1.4~10^{-6}$, for a total of
41 data points.
\label{gendata}}
}

\section{Neural network training strategy}
As described in Ref. \cite{f2ns}, the next step of our strategy is to train
a neural network to each of the replicas of the experimental data.
Artificial neural networks\footnote{
For an introduction to neural networks,
see Ref. \cite{f2ns} and references therein.} (see Fig. \ref{nn}) 
are highly nonlinear
mappings between input and output patterns, defined by its 
parameters, called {\it weights} $\omega_{ij}^{(l)}$ 
and {\it thresholds} $\theta_{i}^{(l)}$.
They provide unbiased robust universal
approximants to incomplete or noisy data, and they
interpolate between data points with the only assumption of
smoothness.
 A neural network
is a suitable way of parametrizing experimental data
since it is a most unbiased prior, and moreover
in combination with the Monte Carlo methods it provides
a faithful estimation of the uncertainties associated to this
parametrization.

\FIGURE[ht]{\epsfig{width=0.89\textwidth,figure=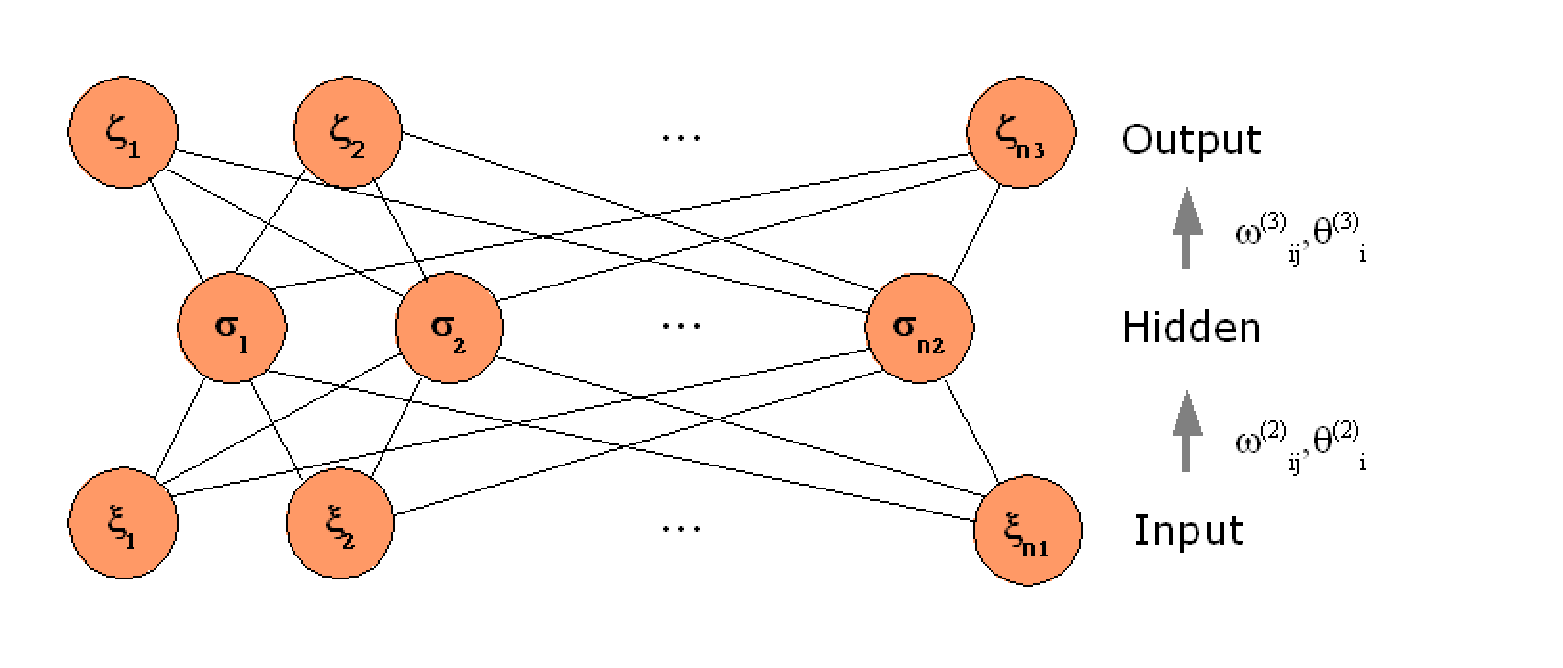} 
        \caption{Schematic representation of
 an artificial neural network.}
        \label{nn}}

In this work a particular class of neural networks called feed-forward
perceptrons are used. For this class of neural networks, 
the relation that gives the values ({\it activation states})
of the i-th neuron in the l-th layer $\xi^{(l)}_i$
depends on the activation states of the
neurons in the previous layer,
\be
 \xi^{(l)}_i=g\lp h_i^{(l)}\rp \ , \quad i=1,\ldots,n_l,
\quad l=1,\ldots,L \ ,
\label{m1}
\ee
\be
\label{m2}
h_i^{(l)}=\sum_{j=1}^{n_{l-1}}\omega_{ij}^{(l)}\xi_{j}^{(l-1)}+
\theta_{i}^{(l)} \ ,
\ee
where $\theta_{i}^{(l)}$ is the {\it activation threshold}
of the neuron, $L$ the total number
of layers of the network and
$n_l$ the number of neurons in each layer. 
The function $g(x)$ is the {\it activation function} of the
neuron, 
which is taken to be a sigmoid
in the inner layers,
\be
g(x)=\frac{1}{1+\exp(-x)} \ ,
\ee
and a linear activation function $g(x)=x$ for the last neuron to increase
the sensitivity of the network.
For illustrative purposes, let us consider a  simple
neural network, which consists on two input neurons
and one output neuron. If $\xi_1^{(1)}$ and $\xi_2^{(2)}$ are the values
of the input neurons, then the value of the output neuron $\xi^{(2)}_1$
will be, from Eqns. \ref{m1} and \ref{m2},
\be
\xi^{(2)}_1=f\lp \xi_1^{(1)},\xi_2^{(2)}\rp =
\lc 1+\exp\lp-\omega_{11}^{(1)}\xi_1^{(1)}
-\omega_{12}^{(1)}\xi_2^{(1)}\rp\rc^{-1} \ .
\ee
From the above explicit example it is clear that a neural network
is a nonlinear function that relates the input values with the
output values.

Therefore, the lepton energy spectrum is parametrized with a neural
network, 
\be 
\lp  \frac{d\Gamma}{dE_l} \rp^{(\net)}(E_l) \, 
\ee 
where $E_l$ is the lepton energy,
 so that if $E_l$ is the input of the neural network,
then $\lp d\Gamma/dE_l\rp^{(\net)}$ is the associated output.

 Training a 
neural network means the determination of
 its parameters (the neuron weights and thresholds)
 to minimize a suitable statistical estimator.
In our case for each replica 
the 
diagonal error will be minimized, defined as
\be
{\chi^2}^{(k)}=\frac{1}{N_{\dat}}
\sum_{i=1}^{N_{\dat}}\frac{\lp M_{i}^{(\art)(k)}-
 M_{i}^{(\net)(k)}\rp^2}{ \sigma^{(\mrexp)2}_{i,\mathrm{tot}}}
\label{chi2diag} \ ,   \qquad k=1,\ldots,N_{\rep} \ ,
\ee
where $M_{i}^{(\net)(k)}$ is the i-th moment as computed from the
k-th neural network, which is  trained on the k-th replica of
the experimental data, and $\sigma^{(\mrexp)}_{i,\mathrm{tot}}$ is the
total error of the i-th data point.

The minimization technique that will be used for the
neural network training is genetic algorithms, a
minimization strategy that has been used in different
high energy physics applications\cite{gathesis} \footnote{
See also Ref. \cite{quevedo} for a recent application
of Genetic Algorithms as the minimization strategy in a 
high energy physics problem.}.  This method
is specially suitable to find the global
minima of highly nonlinear problems, as  the one
that is being discussed in this work.
Other standard deterministic minimization strategies, 
like for example MINUIT,
are not suitable for this problem since the
parameter space is very large. Genetic algorithms
minimization is divided in steps called {\it generations},
so the number of generations of a given neural network training corresponds
to the {\it training length} of the genetic algorithms minimization process.

\FIGURE[ht]
{\epsfig{width=0.45\textwidth,figure=archcomp.ps} 
\caption{Comparison of fits it to the
experimental data with different neural network architectures}
\label{archcomp}}

As has been discussed previously in detail \cite{f2ns}, 
the choice of the
architecture of the neural network (the number of layers and
the number of neurons in each layer) cannot be derived from
general rules and it must be tailored to each specific 
problem. In particular the neural network has to be redundant, that is,
it has to have a larger number of parameters than the
minimum number required to satisfactorily fit the
patterns that have to be learnt, in this case 
experimental data.
However, the architecture cannot be arbitrarily large
because then the training length gets very large. In this case 
one finds
that an acceptable compromise is given by an architecture 1-4-3-1.
A suitable criterion to choose the optimal
architecture is to take the architecture which is next to the
first stable architecture, that is, the first architecture that can
fit the data and gives the same fit that an architecture with one
less neuron. This way one is sure that the neural network
is redundant for the problem that is considered. 
 Fig. \ref{archcomp}  shows a training to the
experimental data with
three different architectures: first 
one observes that 1-2-2-1 is not capable 
to fit
properly the data, but a more complex architecture 1-3-3-1 can fit
this data. Therefore, the architecture 1-4-3-1 is taken
as the reference architecture for the 
parametrization of the lepton energy spectrum.

The training of a neural network does not follow
general rules either, and the optimal minimization strategy must be
determined for each particular problem. 
In the present situation the training strategy 
that is adopted is the following:
there is a single training epoch 
in which the $\chi^{2(k)}$, 
Eq. \ref{chi2diag}, is minimized with 
dynamical stopping of the training of the replicas. 
That is, for each replica,  
 the training is stopped
either when the condition $\chi^{2(k)}\le\chi^2_{\stopp}$
is satisfied or when the maximum number of generations $N_{gen}$ is
reached. One finds that $\chi^2_{\stopp}= 2$ and $N_{\gen}=3000$ are suitable
choices.
On top of that,  the neural network weights are 
initialized 
between $\lc  -\omega_{\mathrm{init}}/2,\omega_{\mathrm{init}}/2\rc$
randomly,
 with
$\omega_{\mathrm{init}}=10$. The rationale
for this choice is that it can be observed that the natural value
for the neural network weights is $\mathcal{O}\lp 10^1\rp$, so 
the fit will  be faster if the initial values for the
neural network weights are of the same order
of magnitude. 

Finally,  in the training of the replicas the so-called
weighted training  will be used for the genetic algorithms minimization.
As it has been shown in Refs. \cite{f2ns,f2nnp}, it is in general 
useful to weight during the training the different experiments 
according to their $\chi^2$, so that more weight is given to those
experiments with a larger value of their $\chi^2$, so that
the final $\chi^2$ is  more homogeneous than in the
unweighted case.
The essential idea of this technique is that the minimized 
$\chi^2_{\mathrm{minim}}$
is given by
\be
\chi^2_{\mathrm{minim}}=\frac{1}{N_{\dat}}\sum_{j=1}^{N_{\mrexp}}
w_{j}N_{\dat,j}\chi^2_j \ ,
\ee
where $N_{\dat,j}$ is the number of data points and $\chi^2_j$
the value of the $\chi^2$, Eq. \ref{chi2diag}, of the
j-th experiment.
One finds after a detailed analysis that the values $w_{\mathrm{Babar}}=0.3$,
$w_{\mathrm{Belle}}=2$ and $w_{\mathrm{Cleo}}=0.5$ are suited to obtain 
a more even $\chi^2_j$ distribution between experiments.

\subsection{Compatibility between experiments}

\FIGURE[ht]
{\epsfig{width=0.60\textwidth,figure=comp1.ps} 
        \caption{$\chi^2$, Eq. \ref{chi2diag}, 
of the different experiments, for a fit to the
experimental data including all experiments.}
        \label{comp1}}

\DOUBLEFIGURE[ht]{comp2.ps,width=0.50\textwidth}{comp3,
width=0.50\textwidth} 
        {$\chi^2$ of the different experiments when 
only the Babar data is fitted.
\label{comp2}}{Same as in Fig. \ref{comp2}, but
now only the Belle data is fitted. 
\label{comp3}}

In global analysis of experimental data which consist of
different experiments, as it is the case now (with Babar, Belle and Cleo),
one has also to address the issue of 
possible inconsistencies between different experiments, that is,
the possibility that a subset of points from 
two experiments in the same region of the parameter space
do not agree with each other within 
experimental errors. This issue
 is of paramount importance in the context of global 
parton distribution fits, see for example \cite{incon,f2nnp}.
In the present application, it can be seen that the three experiments
yield compatible results, as was already known from 
global fits to B decay data. 

This compatibility can be shown in a different
number of ways. For example, training only one experiment and
checking whether or not the other experiments can be predicted, that is,
whether they have a low $\chi^2$ even if they are not incorporated in the 
fit.
In Fig. \ref{comp1} we show a fit to the experimental data
for which all three experiments (Babar, Belle and Cleo)
are incorporated in the fit. One observes that at the
end of the training all experiments satisfy $\chi^2\ll 1$.
In Fig. \ref{comp2} show the results of a fit when only
Babar is incorporated in the fit, and in Fig. \ref{comp3} the
same fit with this time with data from the
Belle experiment only.
Note that when only a single experiment is incorporated in the
fit, like in Figs.  \ref{comp2} and \ref{comp3}, only the
$\chi^2$ of that experiment is expected to decrease, while the total
$\chi^2$ might decrease slower or even grow.

It is observed that the
three experiments are not only compatible but also
complementary.
In particular Cleo is predicted by both Belle and Babar 
(as expected since the kinematical coverage of the
Cleo experiment is included in the other two experiments), while
Belle and Babar cannot predict each other, as a consequence of
the fact that different regions in the parameter space
are covered by the two experiments: only Babar has experimental data
on the $n=0$ moment (partial decay rate), while
only Belle has data at the lowest lepton energy ($E_0=0.4$ GeV).
It will be shown in Section 6 that
also for the correct estimation of the uncertainties the
inclusion of data
from different experiments is crucial.

\subsection{Kinematical constraints}
The lepton spectrum, Eq. \ref{specdef}, has to satisfy three
constraints independently of the dynamics of the process. 
First of all, it vanishes
outside the region where it has kinematical support, in particular
it has to vanish at the kinematical endpoints, $E_l=0$ and
$E_l=E_{\max}$.
Second, the spectrum is a positive definite quantity (since any
integral over it is an observable, a partial branching 
ratio), therefore, it must satisfy a local positivity condition. 

There are several methods  to
introduce kinematical constraints in our
parametrization. 
It has been found that for the present application, the optimal method 
to implement the kinematical constraint that the spectrum
should vanish at the endpoints is to hard-wire them
 into the parametrization, that is,
the lepton energy spectrum parametrized by a neural network will be
given by
\be
\label{gammacon}
\lp \frac{d\Gamma}{dE_l}\rp^{(\net)}(E_l)
=E_l^{n_1}\xi_1^{(L)}(E_l)(E_{\max}-E_l)^{n_2}
\ee
with $n_1,n_2$ positive numbers, and $\xi_1^{(L)}$ is the output
of the neural network for a given value of
$E_l$. The assumption of this functional behavior at the
endpoints of the spectrum introduces no bias since, as will be
shown in Section 6, our results
do not depend on the value of $n_1,n_2$. For the reference training
the values $n_1=1$ and $n_2=1$ have been chosen.

The remaining kinematical constraint, the
positivity constraint, is imposed as a
Lagrange multiplier in the total error. That is, the 
quantity to be minimized, $\chi^2_{\mathrm{tot},\mathrm{min}}$,
 is the sum of two terms,
\be
\label{chi2totcon}
\chi^2_{\mathrm{tot},\mathrm{min}}=\chi^2_{\dat}+\chi^2_{\mathrm{pos}} \ ,
\ee
where the contribution from experimental data $\chi^2_{\dat}$ is 
Eq. \ref{chi2diag}  and the
contribution from the positivity constraint is 
defined as 
\be
\chi^2_{\mathrm{pos}}=\lambda P\lc
\lp \frac{d\Gamma}{dE_l}
\rp^{(\net)} \rc \ ,
\label{chi2cos}
\ee
where the positivity condition is implemented 
in a way that those configurations
in which a region of the spectrum is negative are penalized,
\be
 P\lc
\lp \frac{d\Gamma}{dE_l}
\rp^{(\net)} \rc=-\int_0^{E_{\max}} dE_l \lp \frac{d\Gamma}{dE_l}
\rp^{(\net)}(E_l) \theta \lp -
 \lp \frac{d\Gamma}{dE_l}
\rp^{(\net)}(E_l) \rp,
\ee
since P is zero for a positive spectrum, and leads
to a positive contribution to the total error function, 
Eq. \ref{chi2totcon} if some part of the 
lepton spectrum is negative.
The relative weight $\lambda$ in Eq. \ref{chi2cos} is determined via a
stability analysis, with the
requirement that $\lambda$ is large
enough so that the constraint is verified, but small
enough so that experimental data can still be learned in
an efficient way. It has been found that $\lambda=10^{10}$ satisfies the
above requirements. As  will be proved in the next section,
the implementation of the kinematical constraints plays a essential
role in the parametrization of the lepton spectrum. 

\section{Results}
\label{results}

\FIGURE[ht]{\epsfig{width=0.70\textwidth,figure=specav.ps} 
        \caption{The 1-$\sigma$ uncertainty band for the
lepton energy spectrum, Eq. \ref{specdef},
as parametrized by the Monte Carlo ensemble of neural networks.}
        \label{specav2}}

In this section the results on the parametrization 
of the lepton energy
spectrum are presented. These results
 consist on the sample of trained neural
networks, from which averages and moments can be computed with
the associated uncertainties. The most technical details
of these results for neural network parametrization and the
associated statistical validation can be
found in Appendix A.

\subsection{Lepton energy spectrum}
In figure \ref{specav2}  the  lepton energy spectrum
with uncertainties is represented. For illustration, let us
 recall how the
central value and the spread of this spectrum are
computed from the neural network sample. For the
average one has
\be
\la  
\label{f1}
\lp \frac{d\Gamma}{dE_l}
\rp^{(\net)} \ra (E_l)=\frac{1}{N_{rep}}\sum_{k=1}^{N_{\rep}}
\lp \frac{d\Gamma}{dE_l}
\rp^{(\net)(k)}(E_l) \ ,
\ee
while for the spread the appropriate expression is
\be
\label{f2}
\sigma^{(\net)2}_{d\Gamma/dE_l}(E_l)=
\la \lp \frac{d\Gamma}{dE_l}
\rp^{(\net)2} \ra (E_l)-\lc \la \lp \frac{d\Gamma}{dE_l}
\rp^{(\net)} \ra \rc^2 (E_l) \ .
\ee
In Fig. \ref{specav2}  the $1-\sigma$ envelope
of the lepton spectrum is plotted, where the central value
has been computed with Eq. \ref{f1} and the standard deviation
with Eq \ref{f2}. Note that the error is rather small
for large values of the lepton energy $E_l \ge 1.8$ GeV,
and it grows for smaller values of $E_l$.
 The error bands for the other plots are computed
in the same way.

As discussed before, the sample of trained neural network
reproduce the correlations of the experimental data.
For instance, imagine that one is interested in the
correlation between two moments of the lepton
spectrum, $M_{n_1}(E_{01})$ and $M_{n_2}(E_{02})$,
of arbitrary order and arbitrary lepton energy cut.
With the probability measure of the lepton energy
spectrum constructed in this work, this
correlation $\rho_{12}\equiv \rho\lp n_1,E_{01},n_2,E_{02}\rp$ is given by
\be
\rho_{12}=\frac{
\la M_{n_1}^{(\net)}(E_{01})M_{n_2}^{(\net)}(E_{02}) \ra_{\rep}-
\la M_{n_1}^{(\net)}(E_{01}) \ra_{\rep}\la M_{n_2}^{(\net)}(E_{02})
 \ra_{\rep}}{\sqrt{\la 
M_{n_1}^{(\net)}(E_{01})^2\ra_{\rep}-
\la 
M_{n_1}^{(\net)}(E_{01})\ra_{\rep}^2}
\sqrt{\la 
M_{n_2}^{(\net)}(E_{02})^2\ra_{\rep}-
\la 
M_{n_2}^{(\net)}(E_{02})\ra_{\rep}^2}} \ ,
\ee
where averages over the sample of neural networks
are computed in the standard way, for instance
\be
\la M_{n_1}^{(\net)}(E_{01})M_{n_2}^{(\net)}(E_{02}) \ra_{\rep}=
\frac{1}{N_{\rep}}\sum_{k=1}^{N_{\rep}}
 M_{n_1}^{(\net)(k)}(E_{01})M_{n_2}^{(\net)(k)}(E_{02}) 
\ ,
\ee
and similarly for the remaining averages. This examples clarifies
that fact that not only central values and total
errors from experimental data, showed in Fig. \ref{specav2},
but also correlations are present in the parametrization of the
lepton energy spectrum.

As it has been explained in \cite{f2ns}, it is crucial to
validate the results of the parametrization with suitable 
statistical estimators. In Table \ref{resdata1} 
the most relevant statistical estimators for all the
data points are summarized, 
and in Table \ref{resdata2} one has the same estimators
for the different experiments included in the fit. A more
detailed analysis of these estimators is found in Appendix A.

It has been checked that the large value of $\chi^2$
of the BELLE experiment is not because globally their
data is not properly fitted (as can be seen in the plots),
but that it is only due to two points, $n=2,3$, $E_0=1.5$ GeV
that have an anomalously large $\chi^2$. If those two points are not
included then $\chi^2_{\tot,\mathrm{Belle}}=0.92$. 
These two points are systematically below
Babar data with errors half as small. This is similar as what
happened with the NMC experiment is the proton
structure function fits described in Ref. \cite{f2ns}.

\TABLE[ht]{
\begin{tabular}{|c|ccc|} 
\hline
 $\qquad$ $\qquad$$\qquad$
$\qquad$  &$\qquad$  10  $\qquad$ & $\qquad$ 100  $\qquad$ & 1000\\
\hline
$\chi^2_{\tot}$  & 1.31  & 1.18 & 1.21 \\
$\la \chi^2\ra $   & 2.50      & 2.28 & 2.33  \\
\hline
$\la PE\lc\la M \ra_{\rep}\rc\ra$ &  9\% & 8\% & 8\%\\
$r\lc M \rc$ & 0.999 & 0.999 & 0.999 \\
\hline
$\la PE\lc \sigma^{(\net)}\rc\ra_{\dat}$ &  67\% & 58\% & 45\%\\
$\la \sigma^{(\exp)}\ra_{\dat}$ &  0.00267 & 0.00267 & 0.00267\\
$\la \sigma^{(\net)}\ra_{\dat}$ & 0.00180 & 0.00169 & 0.00187 \\
$r\lc \sigma^{(\net)}\rc$ &  0.77 &  0.85 &  0.86\\
\hline
$\la \rho^{(\exp)}\ra_{\dat}$ &  0.166 & 0.166 &  0.166\\
$\la \rho^{(\net)}\ra_{\dat}$ &  0.320 & 0.245  & 0.324\\
$r\lc \rho^{(\net)}\rc$ & 0.35 & 0.38  & 0.38\\
\hline
$\la \mathrm{cov}^{(\exp)}\ra_{\dat}$ &
$1.4~10^{-6}$ & $1.4~10^{-6}$ & $1.4~10^{-6}$\\
$\la \mathrm{cov}^{(\net)}\ra_{\dat}$ &  $7.8~10^{-7}$ & $6.7~10^{-7}$   &
$1.0~10^{-6}$\\
$r\lc \mathrm{cov}^{(\net)}\rc$ & 0.49  & 0.53  & 0.53 \\
\hline
\end{tabular}
\caption{\small Statistical estimators for the
ensemble of trained neural networks, for 10, 100 and 1000
trained replicas}
\label{resdata1}
}

\TABLE[ht]{
\begin{tabular}{|c|ccc|} 
\hline
  &  Babar  & Belle & Cleo  \\
\hline
$\chi^2_{\tot}$  & 0.42 & 2.06 & 1.22\\
$\la \chi^2\ra $   & 1.67   & 3.13  & 2.21  \\
\hline
$\la PE\lc\la M \ra_{\rep}\rc\ra$ & 2.3\% & 18.1\% & 0.6\%\\
$r\lc M \rc$ & 0.999 & 0.999& 0.999\\
\hline
$\la PE\lc \sigma^{(\net)}\rc\ra_{\dat}$ & 34\% & 44\%& 65\% \\
$\la \sigma^{(\exp)}\ra_{\dat}$ & 0.0023 & 0.0021& 0.0041\\
$\la \sigma^{(\net)}\ra_{\dat}$ & 0.0018& 0.0017 & 0.0022 \\
$r\lc \sigma^{(\net)}\rc$ &  0.94 & 0.89 & 0.83\\
\hline
$\la \rho^{(\exp)}\ra_{\dat}$ & 0.16 & 0.40 & 0.31\\
$\la \rho^{(\net)}\ra_{\dat}$ & 0.15 & 0.28 & 0.51\\
$r\lc \rho^{(\net)}\rc$ & 0.87 & 0.29 & 0.31 \\
\hline
$\la \mathrm{cov}^{(\exp)}\ra_{\dat}$ & $6.9~10^{-6}$& $1.5~10^{-6}$ &
$6.5~10^{-7}$\\
$\la \mathrm{cov}^{(\net)}\ra_{\dat}$ & $2.0~10^{-5}$ &$1.2~10^{-6}$ &
$1.8~10^{-6}$\\
$r\lc \mathrm{cov}^{(\net)}\rc$ &  0.98 & 0.58 & -0.21 \\
\hline
\end{tabular}
\caption{\small Statistical estimators for the
ensemble of trained neural networks, for those experiments
included in the fit. The replica sample consists of 1000 neural networks.}
\label{resdata2}
}

In Figs. \ref{momav0} to \ref{momav3} the computation of the
moments of the lepton energy spectrum 
from our parametrization is compared to the experimental data from
Babar, Belle and Cleo, and good agreement for all the
data points is observed. Note that some of the
experimental data points have not been included in the
training, for the reasons discussed in Section \ref{treatcor},
but nevertheless the lepton energy parametrization is in
good agreement also with 
those data points.

To asses the relevance of the implementation of
kinematical constraints into our neural network 
parametrization of the
lepton spectrum, it is instructive to
 compare   fits with and without the inclusion of kinematical
constraints.
In Fig. \ref{specav_nokincos}
one can observe that when the endpoint
constraint at $E_l=0$ and the positivity constraint are
removed the error becomes very large at small $E_l$. 
This is so because experimental data does not
constrain the value of the lepton spectrum
for low values of $E_l$. 
Note that
the physical value for the spectrum at the endpoint, $\lp d\Gamma
/dE_l\rp (E_l=0)=0$,
is contained within the small-$E_l$ error bars. 

\DOUBLEFIGURE[ht]{momav1.ps, width=.5\textwidth}{momav2.ps,
 width=.5\textwidth}{Comparison of the partial branching
ratio, Eq. \ref{momexp1} obtained from our parametrization
with the experimental measurements, as a function of the lower
cut on the lepton energy $E_0$.  \label{momav0}}{Same as. 
Fig \ref{momav0} for the first moment $M_1$, Eq. \ref{babmom}}

\DOUBLEFIGURE[ht]{momav3.ps, width=.5\textwidth}{momav4.ps,
 width=.5\textwidth}{Same as Fig. \ref{momav0} for the second
moment, $M_2$, Eq. \ref{momexp2}.}{Same as Fig. \ref{momav0} for the third
moment, $M_3$, Eq. \ref{momexp2}. \label{momav3}}

\FIGURE[ht]{\epsfig{width=0.70\textwidth,figure=specav_nokincos.ps} 
        \caption{Lepton energy spectrum when
no kinematic constraints are incorporated in the fit. One
one can see in this case the error at small $E_l$
grows very large and the extrapolation to $E_0=0$ becomes
unreliable.
        \label{specav_nokincos}}}

\DOUBLEFIGURE[ht]{specav_babar.ps,width=0.47\textwidth} 
{specav_ndep.ps,width=0.47\textwidth} 
{Lepton energy spectrum when
only Babar data is incorporated in the fit.
\label{specav_babar} }
       {Comparison of the lepton energy spectrum when
the for different values of the
parameters $n_1$ and $n_2$ \label{specav_ndep}}

To estimate the contribution of the different experiments
to the global fit, it is interesting to compare (see Fig. \ref{specav_babar})  
a fit in which only one  experiment, Babar is incorporated in the
fit. It can be observed that when only the 
Babar data is fitted, the error at small
values of $E_l$ is much larger. This is so because, as discussed 
above, the Belle data, which extends to lower values of $E_l$, is crucial
to determine the low $E_l$ behaviour of the lepton spectrum, together
with the kinematical constraint at $E_l=0$.
Finally,  Fig. \ref{specav_ndep} shows that our results
are independent of the precise choice of $n_1$ and $n_2$ in
Eq. \ref{gammacon}. In particular a fit with the values 
$n_1=1.5$ and $n_2=1.5$ in Eq. \ref{gammacon}
gives the same results as the fit with the reference values, 
$n_1=1$ and $n_2=2$.

With the results described in this section the 
total branching ratio can be computed, even
if experimental information was restricted to a finite 
value of $E_0$. This is possible because
the continuity condition implicit in the neural
network definition together with  the kinematic
constraint  allow for  an accurate
extrapolation from the
experimental data with lowest $E_l=0.4$ GeV to the
kinematic endpoint $E_l=0$. 
Note that this is not true
if the Belle data is not included in the fit,
see Fig. \ref{specav_babar}. 
The result that is obtained for the
partial decay rate,
 computed from the neural network
sample,
\be
\la \mathcal{B}\lp B\to X_c l\nu\rp \ra= 
\la M_0(E_l=0)\ra=\tau_B\frac{1}{N_{\rep}}\sum_{k=1}^{N_{\rep}}
\int_0^{E_{\max}} dE_l \lp \frac{d\Gamma}{dE_l}\rp^{(\net)(k)} \lp E_l
\rp\ ,
\ee
is the total branching ratio,
\be
\mathcal{B}\lp B\to X_c l\nu\rp=\lp 10.8 \pm 0.4\rp~10^{-2} \ ,
\ee
which is to be compared with the 2005 update
for the PDG result \cite{pdg} for the average branching
ratio of neutral and charged B mesons,
\be
\mathcal{B}\lp B\to X_c l\nu\rp_{\mathrm{PDG}}=\lp 10.87 \pm 0.17
\rp~10^{-2} \ ,
\ee
and with the
direct Delphi measurement of the total branching ratio \cite{delphifit},
which is measured without restrictions on the lepton energy,
which yields
\be
\mathcal{B}\lp B\to X_c l\nu\rp_{\mathrm{Delphi}}=
\lp 10.5 \pm 0.2\rp~10^{-2} \ .
\ee
Is is observed that the three results are compatible, even
if our determination is somewhat closer, both in the central
value and in the size of the uncertainty, to the Delphi
measurement.
The small error in our
determination of $\mathcal{B}\lp B\to X_c l\nu\rp$ shows that the
technique discussed in this work can be used also to extrapolate
in a faithful way into regions where there is no
experimental data available.

The results of this section show that from the available
experimental data one can reconstruct the underlying
lepton energy  spectrum with good accuracy.

\subsection{Comparison with theoretical predictions}
As one example of the  applications of the 
present parametrization of the lepton energy spectrum,
in this section our results are 
compared with the
theoretical calculation of Ref. \cite{agru} (AGRU). Their
formalism allows the computation of moments of
different differential distributions from
semileptonic B meson decays, with arbitrary kinematical cuts,
like the lepton energy spectrum in charmed decays
that is analyzed in the present work. 
In particular  their computation of
the lepton energy spectrum will be studied, which they define as
\be
N_k\equiv\frac{1}{\Gamma_{\mathrm{LO}}}\int_{E_0}^{E_{\max}} 
dE_ldq^2dr \tilde{E}_l^k\frac{d^3\Gamma}{dE_ldrdq^2}\ ,
\ee
with $\tilde{E}\equiv E/m_b$, and 
where the leading order partonic semileptonic decay rate 
$\Gamma_{\mathrm{LO}}$ is
given by 
\be
\Gamma_{\mathrm{LO}}=\Gamma_0|V_{cb}|^2 
z_0(\rho) \ ,
\ee
where $\rho\equiv m_c^2/m_b^2$ and the phase space factor is
defined in Eq. \ref{zo}.
These moments can be related to the moments as measured 
experimentally , defined in Eqns. (\ref{momexp1}-
\ref{momexp2}), in a straightforward way, for example for the
first two moments one has
\be
M_0=\tau_B\Gamma_0N_0, \qquad M_1=m_b\frac{N_1}{N_0} \ ,
\ee
and similarly for the other moments.

\DOUBLEFIGURE[ht]{agru1.ps,width=0.47\textwidth}{agru2.ps,
width=0.47\textwidth} 
     {Comparison of the results of Ref. \cite{agru}
on the partial branching ratio Eq. \ref{momexp1} and  the
same quantity computed from our parametrization. 
     \label{agruplot}}{Same as Fig. \ref{agruplot} but for first moment,
Eq. \ref{babmom}. \label{agruplot2}}

\DOUBLEFIGURE[ht]{agru1_err.ps,width=0.47\textwidth}{agru2_err.ps,
width=0.47\textwidth} 
     {Comparison of the results of Ref. \cite{agru}
on the partial branching ratio Eq. \ref{momexp1} at NLO
with associated theoretical uncertainties with  the
same quantity computed from our parametrization. 
     \label{agruplot3}}{Same as Fig. \ref{agruplot3} but for first moment,
Eq. \ref{babmom}. \label{agruplot4}}

\FIGURE[ht]{\epsfig{width=0.60\textwidth,figure=agru3_err.ps} 
        \caption{Comparison of the results of Ref. \cite{agru}
on the fourth moment $\overline{M}_4(E_0)$ at NLO
with associated theoretical uncertainties with  the
same quantity computed from our parametrization.
Note that this moment has not been measured experimentally.
} \label{agru3err}}

 In Figs.  \ref{agruplot} 
and  \ref{agruplot2}\footnote{We thank G. Ridolfi
for providing us with the code used for their
calculations},  the results of \cite{agru} both at leading order (LO)
and at next-to-leading order (NLO) are
compared with the
moments obtained from our parametrization as a function of the lower
cut in the lepton energy $E_0$. Comparing
results at different perturbative orders is interesting
to asses the behaviour of the perturbative expansion.
 One should
take into account in this comparison that the results of
\cite{agru} are purely perturbative, therefore the 
difference between the two results could be an indicator of the
size of the missing nonperturbative corrections. 
Another interesting feature is that while for $M_0$, the partial
branching fraction, the NLO corrections are sizable and
bring the theoretical prediction in better agreement with
the experimental measurement, for $M_1$ (which is the
ratio of two perturbative expansions) the size
of the perturbative corrections is much smaller.

 In Figs.  \ref{agruplot3} 
and  \ref{agruplot4} we show similar results as those of
Figs.  \ref{agruplot} 
and  \ref{agruplot2} but this time with an estimation of the
theoretical uncertainties associated to the predictions
of Ref. \cite{agru}. These theoretical
uncertainties are obtained by varying the
b quark mass $m_b$ 100 MeV above and below
the central value, and similarly for
the strong coupling $\alpha_s(m_b^2)$.
 The known fact that the uncertainties
of theoretical predictions grows for large values
of the cut in the lepton energy $E_0$ is clearly observed
in these results. Note that in all cases comparison
of theoretical predictions with theoretical
measurements can be performed for arbitrary values
of the cut in the lepton energy $E_0$.

On top of that, in Fig. \ref{agru3err} we compare a quantity that has not been
measured, the fourth moment of the spectrum, defined
as
\be
\widetilde{M}_4\lp E_0\rp=\frac{L_4(E_0,0)}{L_0(E_0,0)}=
m_b^4\frac{N_4(E_0)}{N_0(E_0)} \ .
\ee
We observe good agreement for the theoretical
prediction and the experimental data in the region
$ 0.8\le E_0 \le 1.5$ GeV with rather small
uncertainties in both cases. It can also be seen
that for  $E_0 \ge 1.5$ GeV the theoretical uncertainty
for this moment grows while the experimental uncertainty
remains rather small, which implies that theoretical uncertainties
should be reduced at least by a factor of 2 or 3 to
be able to perform quantitative comparisons with
experimental data for moments with large cut in the
lepton energy.
Note therefore that the results in this section imply 
quantitatively how
the uncertainties in theoretical predictions should
be reduced to obtain a meaningful comparison with
experimental data, for example for moments with large cuts
in the lepton energy $E_0$.

A more general comparison with theoretical predictions should include
also the known nonperturbative power corrections up to
order $\mathcal{O}(1/m_b^3)$ to the
expressions for the moments of the spectrum, since in this case
the difference of the theoretical results from our
parametrization would indicate the size of the missing 
unknown corrections,
both perturbative and nonperturbative. A more
detailed study of this point, together with an analysis of possible
violations of local quark-hadron duality, is left for future work. 

The analysis presented in this section is a particular example
of how the technique introduced in this work
allows a more general comparison of theoretical
predictions with experimental data. For example, 
current experiments do not measure the leptonic
moments with $E_0 > 1.5$ GeV, since it is argued
that the corresponding theoretical prediction has
large uncertainties. If in the future this theoretical
error in the computation of leptonic moments
with large values for the cut $E_0$ is reduced, comparison
with experimental results can be straightforward, if
one computes these moments from the neural network
parametrization of the lepton spectrum, which
encodes all the information on available
experimental data.

\section{Determination of $m_b^{1S}$ and $\lambda_1$}

As another
application of our parametrization of the lepton
energy spectrum,
it will be used to determine the
b quark mass $m_b^{1S}$ from the
experimental data using a novel
strategy. To this purpose the
technique of Ref. \cite{bauertrott} will be used, which
consists on
the minimization of the size of higher order corrections
to obtain sets of moments of the
lepton energy spectrum which have reduced theoretical
uncertainty
for the extraction of  nonperturbative parameters like 
$\bar{\Lambda}_{1S}$ and $
\lambda_1$.  
Note that the nonperturbative
parameter $\bar{\Lambda}_{1S}$ relates the spin-averaged
 B meson mass $
\bar{m}_B$
to the 1S scheme b-quark mass $m_b^{1S}$,
\be
\bar{\Lambda}_{1S}\equiv \bar{m}_B-m_b^{1S} \ .
\ee
The 1S scheme b-quark mass
is related to the standard b quark pole mass $m_b^{\mathrm{pole}}$
by a perturbative relation,
\be
m_b^{1S}=m_b^{\mathrm{pole}}\lp 1+\sum_{k=2}^{\infty}C_k\alpha_s\lp
m_b\rp^k\rp  \ .
\ee
The use of heavy quark masses, like the 1S mass or the
MSbar mass, which are not infrared sensitive, is
compulsary to avoid the uncertainties assciated to the
renormalon divergence of some infrared sensitive definitions
of the heavy quark masses, like the pole mass.
The parameter $\lambda_1$ that appears in 
Eq. \ref{gammamb3} is  related to the
the definition of the heavy quark pole masses \cite{gremm} in the
following way,
\be
m_b^{\mathrm{pole}}-m_c^{\mathrm{pole}}=\bar{m}_B-\bar{m}_D+\lambda_1
\frac{\bar{m}_B-\bar{m}_D}{2 \bar{m}_B \bar{m}_D}+\mathcal{O}\lp
\frac{1}{\bar{m}_B^2}\rp \ .
\ee

The moments that minimize the impact of
the higher order nonperturbative corrections are given by
\be
R_1\equiv\frac{
\int_{1.3}^{E_{\max}}E_l^{1.4}
 \frac{d\Gamma}{dE_l}dE_l}{\int_1^{E_{\max}}E_l \frac{d\Gamma}{dE_l}dE_l}\ ,
\ee
and
\be
R_2\equiv\frac{
\int_{1.4}^{E_{\max}}E_l^{1.7}
 \frac{d\Gamma}{dE_l}dE_l}{\int_{0.8}^{E_{\max}}E_l^{1.2} 
\frac{d\Gamma}{dE_l}dE_l} \ .
\ee
The full expression for this moments in terms of
heavy-quark non-perturbative 
parameters can be found in Ref. \cite{bauertrott}, where in terms
or their original notation one has $R_1\equiv R_a^{(1)}$
and $R_2\equiv R_a^{(2)}$.
These leptonic moments $R_1$  and $R_2$ depend on 9 nonperturbative
parameters, up to $\mathcal{O}(1/m_b^{3})$: $\bar{\Lambda}_{1S}$,
$\lambda_1$ and $\lambda_2$, and six matrix elements,
$\rho_1,\rho_2,\tau_1,\tau_2,\tau_3$ and $\tau_4$, that
contribute at order $1/m_b^3$ in the heavy quark
expansion. Present data is 
not capable of a determination of all these matrix elements.
For the $\lambda_2$ parameter we use
\be
\lambda_2\lp m_b\rp=\frac{
m_b^2\Delta m_B-m_c^2\Delta m_D}{2\lp m_b-\kappa m_c\rp} \ ,
\ee
while to 
to asses the contribution of the $\mathcal{O}\lp
1/m_b^3\rp$ parameters, the matrix elements  $\tau_i$ 
are varied between 
$\pm \lp 500~\mathrm{MeV}\rp^3$ (the expected size of this matrix elements),
 $\rho_1$ between 0 and $ 
\lp 500~\mathrm{MeV}\rp^3$ (since from the vacuum saturation
approximation one knows that $\rho_1\ge 0$), 
and for the matrix element $\rho_2$ one uses the relation
 from the power  corrections to the meson mass
splittings \cite{gremm,shapevar},
\be
\rho_2=\tau_2+\tau_4+\frac{m_bm_c \lc \kappa
m_b\Delta m_B-m_c\Delta m_D\rc}{2\lp m_b-\kappa m_c\rp}\  ,
\ee
where we have defined 
\be
\kappa=\lp \frac{\alpha_s(m_c)}{\alpha_s(m_b)}\rp^{3/\beta_0}, \qquad 
\Delta m_Q=m_{Q*}-m_Q \ ,
\ee
and where $\kappa$ account for the scale dependence of the
parameter $\lambda_2$.  

The most relevant feature of these leptonic moments $R_1$ and $R_2$ 
is that they have non-integer powers and to the
best of our knowledge have not been yet
experimentally measured, at least in a publised
form. Therefore,  
 the values of $R_1$ and $R_2$ that will be
used in this analysis are extracted from 
our neural network parametrization of the
lepton spectrum, which allows the
computation of arbitrary moments, together  with their
associated error and correlation. Let us recall that the central values
are determined as
\be
\la R_1^{(\net)}\ra=\frac{1}{N_{\rep}}\sum_{k=1}^{N_{\rep}}
R_1^{(\net)(k)} , \qquad R_1^{(\net)(k)}=
\frac{
\int_{1.3}^{E_{\max}}E_l^{1.4}
\frac{d\Gamma^{(\net)(k)}}{dE_l}(E_l)dE_l}{\int_1^{E_{\max}}
E_l\frac{d\Gamma^{(\net)(k)}}{dE_l} (E_l)dE_l} \ ,
\ee
and similarly for $R_2$,
and the error and the correlation of the moments $R_1$ and 
$R_2$ are computed in the
standard way. The following values for the moments with
the associated errors and their
correlation are obtained,
\be
R_1^{(\net)}=1.017\pm 0.003,  \quad R_2^{(\net)}=0.938\pm 0.004, \quad
\rho_{12}=0.94 \ ,
\ee
that as expected are highly correlated.
Then to determine the nonperturbative parameters $\Lambda_{1S}$
and $\lambda_1$ the associated 
$\chi^2_{\mathrm{fit}}$ is minimized,
\be
\label{chi2fit}
\chi^2_{\mathrm{fit}} = \sum_{i,j=1}^2 \lp R_i^{(\net)}-R_i^{(\tth)}\rp \lp
\mathrm{cov}^{-1}\rp_{ij} 
\lp R_j^{(\net)}-R_j^{(\tth)}\rp \ ,
\ee
where $\mathrm{cov}^{-1}_{ij}$ is the inverse of the covariance matrix
associated to the two moments $R_1^{(\net)}$ and $R_2^{(\net)}$, 
and $R_i^{(\tth)}\lp
\bar{\Lambda}_{1S},\lambda_1\rp$
is the theoretical prediction for these moments as a function of the
two nonperturbative parameters \cite{bauertrott}. 

Once the values of $\bar{\Lambda}_{1S}$ and $\lambda_1$ have
been determined from the minimization of Eq. \ref{chi2fit},
if one uses for the spin averaged B meson mass the
values for the current world average \cite{pdg}
\be
\bar{m}_{B}=\frac{1}{4}\lp m_P+3m_V\rp=\lp 5.3135 \pm 0.0008\rp~\mathrm{GeV} 
\ ,
\ee
then using the extracted value of $\bar{\Lambda}_{1S}$,
\be
\bar{\Lambda}_{1S}=\lp 0.47 \pm 0.14^{\mathrm{exp}}\pm 0.05^{\mathrm{th}} \rp
 ~\mathrm{GeV}\,
\ee
one obtains for the b quark mass $m_b^{1S}$ mass in the 1S
scheme the following
value:
\be
m_b^{1S}=\bar{m}_{B}-\bar{\Lambda}_{1S}=\lp 4.84 \pm 0.14^{\mathrm{exp}}
\pm 0.05^{\mathrm{th}}\rp ~\mathrm{GeV}=\lp 4.84 \pm 0.15^{\mathrm{tot}}
\rp~\mathrm{GeV} \ ,
\ee
From the above results one observes
 that the dominant source of uncertainty is the
experimental uncertainty, that is, the uncertainty associated
to our parametrization of the lepton energy spectrum. This determination of
the b quark mass is consistent with determinations
from other analysis. The b quark mass
has been determined using different techniques, like the sum rule
approach, using either non-relativistic \cite{hoang,hoang2,
benekesum,pinedasumrule}
or relativistic \cite{kuhnbmass,corcella} sum rules, 
 global fits of moments of
differential distributions in
 B decays, \cite{bauerglobalfit,delphifit,fitskin},
 the renormalon analysis of Ref. \cite{pineda}, and several other
methods related to heavy-quarkonium physics \cite{quarkonium1,
quarkonium2}
(see \cite{quarkonium3} for a review). To
compare our results with some of the above references, it is 
useful to 
 relate the $m^{1S}_b$ mass to the MS-bar $\bar{m}_b \lp \bar{m}_b\rp$
mass \cite{hoang2,hoangfull}, and once the conversion
is performed\footnote{We thank
Andre Hoang for pointing us the appropiate
references to perform the mass sheme conversion.} the value
\be
\bar{m}_b\lp \bar{m}_b \rp=\lp 4.31 \pm 0.15^{\tot} \rp~\mathrm{GeV} \ ,
\ee
is obtained,
where we have used $\alpha_s(M_Z^2)=0.1182$ and included 
perturbative corrections up to two loops.
It turns out that our determination of 
$m^{1S}_b$ is not competitive with respect
to other determinations since only two moments,
$R_1$ and $R_2$ are used to constrain the
nonperturbative parameters in the fit.
Note therefore that the relatively large error in the extraction of
$\bar{m}_b\lp \bar{m}_b\rp$ are not due to large uncertainties in our
parametrization of the lepton energy spectrum, which are the
same than experimental data, but rather from the use of
a reduced set of moments in the fit. Note also that the
motivation to perform this determination of
$\bar{m}_b\lp \bar{m}_b\rp$ is to show how the
neural network parametrization constructed in this
work allows a more general comparison of experimental data with
theoretical predictions, in this case allowing to use
moments with fractional index with their errors and
correlations, which have not been measured directly, at least
in a published form.
The inclusion of additional moments  would therefore
constrain more the nonperturbative parameters and
reduce the experimental uncertainty associated to them.

For the nonperturbative parameter $\lambda_1$  the
following value is obtained
\be
\lambda_{1}=\lp -0.16 \pm 0.14^{\mathrm{exp}}  \pm 0.05^{\mathrm{th}}\rp~
\mathrm{GeV^2} =\lp -0.16 \pm 0.15^{\mathrm{tot}}\rp ~\mathrm{GeV^2} 
\ .
\ee
As in the determination of $\bar{\Lambda}_{1S}$ 
it can be seen that the theoretical uncertainties are
smaller than the experimental ones, which are the dominant ones.
Our result for the parameter $\lambda_1$ is consistent with other 
extractions in the context of  global fits of
B decay data \cite{bauerglobalfit,delphifit}, but again not competitive
due to the large experimental uncertainties.

In summary, a determination of $m^{1S}_b$ and  $\lambda_{1}$ has
been obtained from our 
neural network parametrization in a way that was not directly possible
from the available experimental data.
However, it turns out that the experimental uncertainties in the
present determination do not allow these results to be
competitive with those from other determinations
using different techniques, even if in this
approach the theoretical
uncertainties where minimized.

\section{Conclusions and outlook}
This works presents a determination of the probability density in the
space of the lepton energy spectra from semileptonic B meson decays, based
on the latest available data from the Babar, Belle and
Cleo collaborations, that makes use of
 a combination of Monte Carlo techniques
and neural networks with results in an unbiased parametrization
with faithful estimation of the uncertainties.
In addition, this work shows the implementation of
a  well definite strategy to reconstruct 
functions with uncertainties
 when the only available experimental information comes 
through convolutions of these functions. Moreover, in our
formalism the implementation of 
arbitrary theoretical constraints
can be done in a consistent and
unbiased way.

As a byproduct of our analysis, with 
our parametrization of the lepton spectrum,
 the nonperturbative parameters $\bar{\Lambda}_{1S}$
and $\lambda_1$ have been extracted
 in a way that minimizes the theoretical uncertainties.
For the b quark mass in the $1S$ scheme the
result $
m_b^{1S}=\bar{m}_{B}-\bar{\Lambda}_{1S}=\lp 4.84 \pm 0.16^{\mathrm{exp}}
\pm 0.05^{\mathrm{th}}\rp ~\mathrm{GeV}$
has been obtained.
Although this application demonstrates  the flexibility of
our approach to allow a more general comparison of data with theoretical
predictions, it turns out that 
the use of a reduced set of non-integer moments
implies that uncertainties are rather large
to make this determination competitive with those
from global fits of moments of B meson
decay distributions \cite{bauerglobalfit,delphifit,fitskin}, which
include additional moments. A reevaluation of 
$m_b^{1S}$ from our neural network parametrization of the
lepton energy spectrum with a larger set of moments will
be studied in a following work. 

The number of possible applications of this strategy to other 
problems in B physics is rather large. For example, the 
inclusion of hadronic moments would allow
a parametrization of the double differential
decay rate
\be
\frac{d^2\Gamma}{dE_ldr}(E_l,r)
\equiv \int dq^2 \frac{d^3\Gamma}{dq^2dE_ldr} \ ,
\ee
in terms of a neural network with
two inputs. From this two-dimensional
spectrum the hadronic invariant mass moments,
as have been recently measured by Babar \cite{babarhadmom} and Belle
\cite{bellehadmom},
would be computed as
\be
H_n(E_0,\mu)\equiv \int_{E_0}^{E_{\max}}dE_l \int_0^{M_B/2}
dr(r-\mu)^n \frac{d^2\Gamma}{dE_ldr}(E_l,r) \ .
\ee 
In this case both the training and
the implementation of the kinematical constraints is
rather more complicated, since one has to
parametrize a two-dimensional surface.

The charmless decay channel, $B\to X_u l\nu$
would be also interesting to analyze, 
since it has received a lot of theoretical attention recently, 
specially in the context of effective field theories, see
 for example Refs. \cite{neubertvub,ossola,charmlesstheory,charmlessbabar,
vubbabarpaper,gardicharmless}
and references therein. However, this mode
 is more challenging to measure due to large backgrounds.
Another interesting application would be to 
estimate with our technique the issue of the
parametrization dependence of the form factor of the $B\to \pi l\nu$
exclusive channel,
as discussed in Ref. \cite{becher}.

A  process that is closely related to the
semileptonic decays is the
analysis of the photon energy spectrum in $B\to X_s\gamma$ 
decays \cite{anatomy,gardi}. This process has also been
recently measured with good accuracy at the
B factories, by Babar \cite{bxgbabar} and
by Belle \cite{bxgbelle}. The strategy to be followed in this
process would be very similar to that of the present work, since
the experimental information has the same form.

Finally one can use the neural network strategy 
to construct a parametrization of the shape function $S(k)$ 
of the B meson, a universal characteristic
of the B meson that governs inclusive decay spectra
in processes with massless partons in the
final state, as extracted from the
$B\to X_s\gamma$ and $B\to X_ul\nu$ decay modes.
 In this case there exist more theoretical
information on its shape. For example,
at tree level  its moments 
\be
A_n\equiv \int dk k^n S(k)
\ee
have to satisfy $A_0=1,A_1=0$ and $A_2=\mu_{\pi}^2/3$.
At higher orders these relations are theoretically more
controversial \cite{neubertfactorization,shapemanohar}.
Since the uncertainty from the extraction of $S(k)$
 is the dominant source of theoretical 
uncertainty in some CKM matrix
elements extraction, it would therefore be interesting to 
estimate again this uncertainty with the technique presented in this
work, since in the current approach \cite{shape2} 
the shape function uncertainties are estimated in
a rather crude way, with a combination of different functional
forms compatible with the theoretical constraints.
The application of the techniques introduced in this work
to obtain an unbiased parametrization of the B meson shape function
with a faithful estimation of its uncertainties from
experimental data will be discussed in a forthcoming
publication \cite{shapefun}.

The set of trained neural networks that represents the probability measure
in the space of differential lepton energy spectra, together with
the driver program and a user manual are available from the author\footnote{
{\tt joanrojo@ecm.ub.es}}.

\acknowledgments
I want to thank the members of the NNPDF Collaboration, Luigi del Debbio, 
Stefano Forte, Jos\'e Ignacio Latorre and Andrea Piccione, since a sizeable
fraction of this work is related to an upcoming
publication \cite{nnqns} in collaboration with them.
I want also to thank Thomas Becher, Einan Gardi, Antonio Pineda and Giovanni
Ridolfi
 for useful discussions and encouragement, Peter Krizan, Giulio 
D'Agostini and Thosten Brand for help
with the experimental data, and Andre Hoang for useful remarks
about the b quark mass scheme conversion.

\appendix

\section{Details of the neural network training}

In this Appendix a more detailed statistical analysis of the
parametrization of the lepton energy spectrum is performed.
To understand better the process of
neural network training it is interesting to
 analyze the evolution of the different statistical estimators,
as defined in Appendix B,
with respect to the number of {\it generations}, that is,
with respect of the training length.
In fig. \ref{averr}  the
evolution of $\chi^2_{\tot}$ and of 
$\la \chi^2\ra$ computed from the
trained replica sample can be observed. Note that at the end
of the training $\chi^{2}_{\tot}\sim 1$ and $\la \chi^2\ra \sim 2$,
as expected. Note also that the fit has reached convergence with the
$\chi^2_{\tot}$ profile is very flat for a large number of generations. 

This can be repeated for other
statistical estimators, like for example
the average spread of the data points
 as computed from the neural network
ensemble, Fig. \ref{averr2}, defined
in Appendix \ref{estimators},
which is to be compared with the same quantity computed from
the experimental data, $\sigma_i^{(\mrexp)}$. 
The fact that one has error reduction, as has been explained in
\cite{f2ns},
is the sign that the network has found an underlying law
to the experimental data, in this case the lepton
energy spectrum.

Other relevant estimator is the so-called 
 scatter correlations of the spread of the points
(see Fig. \ref{averr3}).
The scatter correlation indicates the size of the spread of data
around a straight line. Specifically $r\lc \sigma^{(\net)}\rc=1$
implies that $\la \sigma_i^{(\net)}\ra$ is proportional
to $\sigma_i^{(\mrexp)}$.
One can define similarly a scatter correlation for the
net correlation $\rho_{ij}^{(\net)}$, also represented in Fig. 
\ref{averr3} for the Babar experimental data. One observes that when
the training ends both values of $r$ are close to 1, a sign
that errors and correlations are  faithfully reproduced.

Another relevant estimator of the goodness of the fit is the distribution 
of both $\chi^{2(k)}$ and of the training lengths over the replica sample,
figures \ref{hist}, and \ref{histtl}. The distribution of $\chi^{2(k)}$ over
the replica sample should be rather peaked around $\la \chi^2\ra$, because
 the opposite case would mean that the averaged result is obtained 
as a combination
very good fits with very bad fits (in the sense of fits with very large 
$\chi^2$). It can be observed in Figure \ref{hist} that indeed our
distribution is very peaked around the average.  On the other
hand, the distribution
of training lenghts, Fig. \ref{histtl},  has to be smooth and it cannot
be peaked at $N_{\gen}$, because if a too large fraction of the
nets never reach the condition $\chi^{2(k)}\le \chi^2_{\mathrm{stop}}$
then effectively one is stopping the training after a fixed number
of generations regardless of the value of the $\chi^{2(k)}$
of the trained replica.
It can be seen that only $\sim 20\%$ of the
trained replicas do not reach $\chi^2_{\stopp}$, which is an acceptable
fraction.

\FIGURE[ht]{\epsfig{width=0.60\textwidth,figure=chi2tot.ps} 
        \caption{Total $\chi^{2}_{\tot}$, Eq. \ref{chi2tot} of the
replica sample, compared 
with average error, $\la \chi^2\ra$, Eq. \ref{averrdiag}. 
}
        \label{averr}}

\DOUBLEFIGURE[ht]{est_averr.ps,width=0.50\textwidth}{est_scorr.ps,
width=0.50\textwidth} 
        {Average error of the data points
as computed from the neural network sample, Eq. \ref{var},
as compared with the experimental value.
\label{averr2}}{The scatter correlations,
	  Eq. \ref{sccnets} as a function of the lenght
of the training, for the
Babar experimental data. \label{averr3}}

\DOUBLEFIGURE[ht]{hist.ps,width=0.50\textwidth}{histtl.ps,
width=0.47\textwidth}{Distribution 
of $\chi^2$ over the sample
	of trained replicas. \label{hist}}{Distribution 
of training lenghts over the sample
	of trained replicas. \label{histtl}}

\section{Statistical estimators}
\label{estimators}
In this appendix the statistical estimators that are
used to asses the quality of both the Monte Carlo replica generation and
the neural network training are described.
The superscripts $(\dat)$, $(\art)$ and $(\net)$ refer respectively to
the original data, to the $N_{\rep}$ Monte Carlo replicas of the data,
and to the  $N_{\rep}$  neural networks.
The subscripts $\rep$ and $\dat$ refer respectively to whether averages 
are taken  by summing over all replicas or over all data.

\begin{itemize}
\item{\bf Replica averages}
\begin{itemize}
\item Average over the number of replicas for each experimental point
  $i$
\be
\la
 M_i^{(\art)}\ra_{\rep}=\frac{1}{N_{\rep}}\sum_{k=1}^{N_{\rep}}
M_i^{(\art)(k)}\ .
\ee 
\item Associated variance
\be
\label{var}
\sigma_i^{(\art)}=\sqrt{\la\lp M_i^{(\art)}\rp^2\ra_{\rep}-
\la M_i^{(\art)}\ra^2_{\rep}} \ .
\ee
\item Associated covariance
\be
\label{ro}
\rho_{ij}^{(\art)}=\frac{\la M_i^{(\art)}M_j^{(\art)}\ra_{\rep}-
\la M_i^{(\art)}\ra_{\rep}\la M_j^{(\art)}\ra_{\rep}}{\sigma_i^{(\art)}
\sigma_j^{(\art)}} \ .
\ee
\be
\label{cov}
\mathrm{cov}_{ij}^{(\art)}=\rho_{ij}^{(\art)}\sigma_i^{(\art)}
\sigma_j^{(\art)} \ .
\ee
\item Percentage error on central values 
over the
  $N_{\dat}$ data points.
\be
\la PE\lc\la M^{(\art)}\ra_{\rep}\rc\ra_{\dat}=
\frac{1}{N_{\dat}}\sum_{i=1}^{N_{\dat}}\lc\frac{ \la M_i^{(\art)}\ra_{\rep}-
M_i^{(\mrexp)}}{M_i^{(\mrexp)}}\rc \ .
\ee
We define analogously $\la PE\lc\la
\sigma^{(\art)}\ra_{\rep}\rc\ra_{\dat}$. 
\item Scatter correlation:
\be
r\lc M^{(\art)}\rc=\frac{\la M^{(\mrexp)}\la M^{(\art)}
\ra_{\rep}\ra_{\dat}-\la M^{(\mrexp)}\ra_{\dat}\la\la M^{(\art)}
\ra_{\rep}\ra_{\dat}}{\sigma_s^{(\mrexp)}\sigma_s^{(\art)}} \,
\ee
where the scatter variances are defined as
\be
\sigma_s^{(\mrexp)}=\sqrt{\la \lp M^{(\exp)}\rp^2\ra_{\dat}-
\lp \la  M^{(\exp)}\ra_{\dat}\rp^2} \,
\ee
\be
\sigma_s^{(\art)}=\sqrt{\la \lp \la M^{(\art)}\ra_{\rep}\rp^2\ra_{\dat}-
\lp \la  \la M^{(\art)}\ra_{\rep} \ra_{\dat}\rp^2} \ .
\ee 
We define analogously $r\lc\sigma^{(\art)}\rc$, $r\lc\rho^{(\art)}\rc$
and  $r\lc\mathrm{cov}^{(\art)}\rc$. Note that the scatter correlation
and
scatter variance are not related to the variance and correlation
Eqs. \ref{var}-\ref{cov}.
\item Average variance:
\be
\la \sigma^{(\art)}\ra_{\dat}=\frac{1}{N_{\dat}}
\sum_{i=1}^{N_{\dat}}\sigma_i^{(\art)} \ .
\label{avvar}
\ee
We  define analogously $\la\rho^{(\art)}\ra_{\dat}$ and
$\la\mathrm{cov}^{(\art)}\ra_{\dat}$,  as well as the
corresponding experimental quantities.
\end{itemize}
\item{\bf Neural network averages}
\begin{itemize}
\item Mean variance and percentage error on central values over the
  $N_{\dat}$ data points.
\be
\la PE\lc\la M^{(\net)}\ra_{\rep}\rc\ra_{\dat}=
\frac{1}{N_{\dat}}\sum_{i=1}^{N_{\dat}}\lc\frac{ \la M_i^{(\net)}\ra_{\rep}-
M_i^{(\mrexp)}}{M_i^{(\mrexp)}}\rc \ .
\label{penets}
\ee
\item
We define analogously percentage errors on the correlation and
covariance. 
\item Scatter correlation
\be
r\lc M^{(\net)}\rc=\frac{\la M^{(\mrexp)}\la M^{(\net)}
\ra_{\rep}\ra_{\dat}-\la M^{(\mrexp)}\ra_{\dat}\la\la M^{(\net)}
\ra_{\rep}\ra_{\dat}}{\sigma_s^{(\mrexp)}\sigma_s^{(\net)}} \ .
\label{sccnets}
\ee 
We define analogously $\la\rho^{(\net)}\ra_{\dat}$ and
$\la\mathrm{cov}^{(\net)}\ra_{\dat}$.

\end{itemize}
\end{itemize}

On top of these, one has  also to
define the estimators that measure the global quality
of the fit, namely the total error
\be
\label{chi2tot}
 \chi^{2}_{\tot}=
\frac{1}{N_{\dat}}\sum_{i=1}^{N_{\dat}}\frac{\lp M_i^{(\exp)}-
\la M_i^{(\net)}\ra_{\rep}\rp^2}{\sigma_{\tot,i}^2} \ ,
\ee and the average error over the
replica sample,
\be
\label{averrdiag}
\la \chi^2\ra=\frac{1}{N_{\rep}}\sum_{i=k}^{N_{\rep}}
\frac{1}{N_{\dat}}\sum_{i=1}^{N_{\dat}}\frac{\lp M_i^{(\art)(k)}-
M_i^{(\net)(k))}\rp^2}{\sigma_{\tot,i}^2} \ .
\ee
On general grounds \cite{f2ns} one expects the relation 
$\la \chi^2 \ra\sim \chi^2_{\tot}+1$ to hold, and indeed this
is the case as can be seen in Tables 4 and 5.

\providecommand{\href}[2]{#2}\begingroup\raggedright\endgroup


\begin{thebibliography}{10}

\bibitem{superB}
Hewett, J. {\em et~al.}, {\it The discovery potential of a super b
  factory.},
  \href{http://xxx.lanl.gov/abs/hep-ph/0503261}{{\tt hep-ph/0503261}}.

\bibitem{gambinomom}
P.~Gambino and N.~Uraltsev, {\it Moments of semileptonic b decay distributions
  in the $1/m_b$ expansion},  {\em Eur. Phys. J.} {\bf C34} (2004) 181--189,
  [\href{http://xxx.lanl.gov/abs/hep-ph/0401063}{{\tt hep-ph/0401063}}].

\bibitem{trott}
M.~Trott, {\it Improving extractions of $V_{cb}$ and the b quark mass from
  semileptonic inclusive b decay},  {\em Int. J. Mod. Phys.} {\bf A19} (2004)
  5493--5500, [\href{http://xxx.lanl.gov/abs/hep-ph/0409186}{{\tt
  hep-ph/0409186}}].

\bibitem{gambinorev}
P.~Gambino, {\it Semileptonic and radiative b decays circa 2005},
  \href{http://xxx.lanl.gov/abs/hep-ph/0510085}{{\tt hep-ph/0510085}}.

\bibitem{ope1}
J.~Chay, H.~Georgi, and B.~Grinstein, {\it Lepton energy distributions in heavy
  meson decays from qcd},  {\em Phys. Lett.} {\bf B247} (1990) 399--405.

\bibitem{ope2}
I.~I.~Y. Bigi, M.~A. Shifman, N.~G. Uraltsev, and A.~I. Vainshtein, {\it QCD
  predictions for lepton spectra in inclusive heavy flavor decays},  {\em Phys.
  Rev. Lett.} {\bf 71} (1993) 496--499,
  [\href{http://xxx.lanl.gov/abs/hep-ph/9304225}{{\tt hep-ph/9304225}}].

\bibitem{f2ns}
S.~Forte, L.~Garrido, J.~I. Latorre, and A.~Piccione, {\it Neural network
  parametrization of deep-inelastic structure functions},  {\em JHEP} {\bf 05}
  (2002) 062, [\href{http://xxx.lanl.gov/abs/hep-ph/0204232}{{\tt
  hep-ph/0204232}}].

\bibitem{f2nnp}
{\bf NNPDF} Collaboration, L.~Del~Debbio, S.~Forte, J.~I. Latorre, A.~Piccione,
  and J.~Rojo, {\it Unbiased determination of the proton structure function
  $F_2^p$ with faithful uncertainty estimation},
  \href{http://xxx.lanl.gov/abs/hep-ph/0501067}{{\tt hep-ph/0501067}}.

\bibitem{tau}
J.~Rojo and J.~I. Latorre, {\it Neural network parametrization of spectral
  functions from hadronic tau decays and determination of qcd vacuum
  condensates},  {\em JHEP} {\bf 01} (2004) 055,
  [\href{http://xxx.lanl.gov/abs/hep-ph/0401047}{{\tt hep-ph/0401047}}].

\bibitem{nnqns}
{\bf NNPDF} Collaboration, L.~Del~Debbio, S.~Forte, J.~I. Latorre, A.~Piccione,
  and J.~Rojo, {\it The neural network approach to parton fitting: The
  nonsinglet case.}

\bibitem{pdf}
{\bf NNPDF} Collaboration, J.~Rojo, L.~Del~Debbio, S.~Forte, J.~I. Latorre, and
  A.~Piccione, {\it The neural network approach to parton fitting},
  \href{http://xxx.lanl.gov/abs/hep-ph/0505044}{{\tt hep-ph/0505044}}.


\bibitem{shapefun}
G.~Ossola, A.~Piccione, and J.~Rojo, {\it Neural network parametrization of the
  B meson shape function},  {\em In preparation}.

\bibitem{heralhc}
M.~Dittmar {\em et~al.}, {\it Parton distributions: Summary report for the HERA
  - LHC workshop},  \href{http://xxx.lanl.gov/abs/hep-ph/0511119}{{\tt
  hep-ph/0511119}}.

\bibitem{bauertrott}
C.~W. Bauer and M.~Trott, {\it Reducing theoretical uncertainties in $m_b$ and
  $\lambda_1$},  {\em Phys. Rev.} {\bf D67} (2003) 014021,
  [\href{http://xxx.lanl.gov/abs/hep-ph/0205039}{{\tt hep-ph/0205039}}].

\bibitem{trottmom}
M.~Trott, {\it Improving extractions of $V_{cb}$ and $m_b$ from the hadronic
  invariant mass moments of semileptonic inclusive B decay},  {\em Phys. Rev.}
  {\bf D70} (2004) 073003, [\href{http://xxx.lanl.gov/abs/hep-ph/0402120}{{\tt
  hep-ph/0402120}}].

\bibitem{agru}
V.~Aquila, P.~Gambino, G.~Ridolfi, and N.~Uraltsev, {\it Perturbative
  corrections to semileptonic b decay distributions},  {\em Nucl. Phys.} {\bf
  B719} (2005) 77--102, [\href{http://xxx.lanl.gov/abs/hep-ph/0503083}{{\tt
  hep-ph/0503083}}].


\bibitem{kuhn}
M.~Jezabek and J.~H. Kuhn, {\it Lepton spectra from heavy quark decay},  {\em
  Nucl. Phys.} {\bf B320} (1989) 20.

\bibitem{blm}
M.~Gremm and I.~Stewart, {\it Order alpha(s)**2 beta(0) correction to the
  charged lepton spectrum in b ---> c l anti-nu/l decays},  {\em Phys. Rev.}
  {\bf D55} (1997) 1226--1232,
  [\href{http://xxx.lanl.gov/abs/hep-ph/9609341}{{\tt hep-ph/9609341}}].


\bibitem{manohar}
A.~V. Manohar and M.~B. Wise, {\it Inclusive semileptonic b and polarized
  $\Lambda_b$ decays from QCD},  {\em Phys. Rev.} {\bf D49} (1994) 1310--1329,
  [\href{http://xxx.lanl.gov/abs/hep-ph/9308246}{{\tt hep-ph/9308246}}].

\bibitem{gremm}
M.~Gremm and A.~Kapustin, {\it Order $1/m_b^3$ corrections to inclusive
  semileptonic B decay},  {\em Phys. Rev.} {\bf D55} (1997) 6924--6932,
  [\href{http://xxx.lanl.gov/abs/hep-ph/9603448}{{\tt hep-ph/9603448}}].

\bibitem{bigitotal}
D.~Benson, I.~I. Bigi, T.~Mannel, and N.~Uraltsev, {\it Imprecated, yet
  impeccable: On the theoretical evaluation of $\Gamma (B \to X_c l \nu)$},
  {\em Nucl. Phys.} {\bf B665} (2003) 367--401,
  [\href{http://xxx.lanl.gov/abs/hep-ph/0302262}{{\tt hep-ph/0302262}}].

\bibitem{Aubert:2004td}
{\bf BABAR} Collaboration, B.~Aubert {\em et~al.}, {\it Measurement of the
  electron energy spectrum and its moments in inclusive $B\to X e \nu$ decays},
   {\em Phys. Rev.} {\bf D69} (2004) 111104,
  [\href{http://xxx.lanl.gov/abs/hep-ex/0403030}{{\tt hep-ex/0403030}}].

\bibitem{hfag}
H.~F.~A. Group {\em et~al.}, {\it Averages of B hadron properties at the end of
  2005},  \href{http://xxx.lanl.gov/abs/hep-ex/0603003}{{\tt hep-ex/0603003}}.

\bibitem{elmombelle}
{\bf Belle} Collaboration, {\it Moments of the electron energy spectrum in $B
  \to X_c l \nu$ decays at belle},
  \href{http://xxx.lanl.gov/abs/hep-ex/0508056}{{\tt hep-ex/0508056}}.

\bibitem{cleoleptonmom}
{\bf CLEO} Collaboration, A.~H. Mahmood {\em et~al.}, {\it Measurement of the
  B-meson inclusive semileptonic branching fraction and electron energy
  moments},  {\em Phys. Rev.} {\bf D70} (2004) 032003,
  [\href{http://xxx.lanl.gov/abs/hep-ex/0403053}{{\tt hep-ex/0403053}}].

\bibitem{cdfhadmom}
{\bf CDF} Collaboration, D.~Acosta {\em et~al.}, {\it Measurement of the
  moments of the hadronic invariant mass distribution in semileptonic B
  decays},  {\em Phys. Rev.} {\bf D71} (2005) 051103,
  [\href{http://xxx.lanl.gov/abs/hep-ex/0502003}{{\tt hep-ex/0502003}}].

\bibitem{fitbabar}
{\bf BABAR} Collaboration, B.~Aubert {\em et~al.}, {\it Determination of the
  branching fraction for $B\to X_c l\nu_l$ decays and of V$_{cb}$ from hadronic
  mass and lepton energy moments},  {\em Phys. Rev. Lett.} {\bf 93} (2004)
  011803, [\href{http://xxx.lanl.gov/abs/hep-ex/0404017}{{\tt
  hep-ex/0404017}}].

\bibitem{fitskin}
O.~Buchmueller and H.~Flaecher, {\it Fits to moment measurements from $B\to X_c
  l\nu$ and $B \to X_s \gamma$ decays using heavy quark expansions in the
  kinetic scheme},  \href{http://xxx.lanl.gov/abs/hep-ph/0507253}{{\tt
  hep-ph/0507253}}.

\bibitem{gathesis}
Y.~A. Kanev, {\it Application of neural networks and genetic algorithms in
  high-energy physics}, PhD. Thesis, UMI-99-05968.

\bibitem{quevedo}
B.~C. Allanach, D.~Grellscheid, and F.~Quevedo, {\it Genetic algorithms and
  experimental discrimination of susy models},  {\em JHEP} {\bf 07} (2004) 069,
  [\href{http://xxx.lanl.gov/abs/hep-ph/0406277}{{\tt hep-ph/0406277}}].

\bibitem{incon}
J.~C. Collins and J.~Pumplin, {\it Tests of goodness of fit to multiple data
  sets},  \href{http://xxx.lanl.gov/abs/hep-ph/0105207}{{\tt hep-ph/0105207}}.

\bibitem{pdg}
{\bf Particle Data Group} Collaboration, S.~Eidelman {\em et~al.}, {\it Review
  of particle physics},  {\em Phys. Lett.} {\bf B592} (2004) 1.

\bibitem{delphifit}
{\bf DELPHI} Collaboration, J.~, Abdallah {\em et~al.}, {\it Determination of
  heavy quark non-perturbative parameters from spectral moments in semileptonic
  b decays},  \href{http://xxx.lanl.gov/abs/hep-ex/0510024}{{\tt
  hep-ex/0510024}}.

\bibitem{hoang}
A.~H. Hoang, {\it Bottom quark mass from upsilon mesons},  {\em Phys. Rev.}
  {\bf D59} (1999) 014039, [\href{http://xxx.lanl.gov/abs/hep-ph/9803454}{{\tt
  hep-ph/9803454}}].

\bibitem{hoang2}
A.~H. Hoang, {\it 1S and MSbar bottom quark masses from upsilon sum rules},  {\em Phys. Rev.}
  {\bf D61} (2000) 034005, [\href{http://xxx.lanl.gov/abs/hep-ph/9905550}{{\tt
  hep-ph/9905550}}].

\bibitem{shapevar}
C.~W. Bauer, Z.~Ligeti, M.~Luke, and A.~V. Manohar, {\it B decay shape
  variables and the precision determination of $|V_{cb}|$ 
and $m_b$},  {\em Phys.
  Rev.} {\bf D67} (2003) 054012,
  [\href{http://xxx.lanl.gov/abs/hep-ph/0210027}{{\tt hep-ph/0210027}}].

\bibitem{benekesum}
M.~Beneke and A.~Signer, {\it The bottom MSbar quark mass from sum rules at
  next-to- next-to-leading order},  {\em Phys. Lett.} {\bf B471} (1999)
  233--243, [\href{http://xxx.lanl.gov/abs/hep-ph/9906475}{{\tt
  hep-ph/9906475}}].

\bibitem{pinedasumrule}
A.~Pineda and A.~Signer, {\it Renormalization group improved sum rule analysis
  for the bottom quark mass},
  \href{http://xxx.lanl.gov/abs/hep-ph/0601185}{{\tt hep-ph/0601185}}.


\bibitem{kuhnbmass}
J.~H. Kuhn and M.~Steinhauser, {\it Determination of $\alpha_s$ and heavy quark
  masses from recent measurements of $R(s)$},  {\em Nucl. Phys.} {\bf B619}
  (2001) 588--602, [\href{http://xxx.lanl.gov/abs/hep-ph/0109084}{{\tt
  hep-ph/0109084}}].

\bibitem{corcella}
G.~Corcella and A.~H. Hoang, {\it Uncertainties in the MSbar bottom quark mass
  from relativistic sum rules},  {\em Phys. Lett.} {\bf B554} (2003) 133--140,
  [\href{http://xxx.lanl.gov/abs/hep-ph/0212297}{{\tt hep-ph/0212297}}].

\bibitem{bauerglobalfit}
C.~W. Bauer, Z.~Ligeti, M.~Luke, A.~V. Manohar, and M.~Trott, {\it Global
  analysis of inclusive B decays},  {\em Phys. Rev.} {\bf D70} (2004) 094017,
  [\href{http://xxx.lanl.gov/abs/hep-ph/0408002}{{\tt hep-ph/0408002}}].

\bibitem{pineda}
A.~Pineda, {\it Determination of the bottom quark mass from the $\Upsilon(1S)$
  system},  {\em JHEP} {\bf 06} (2001) 022,
  [\href{http://xxx.lanl.gov/abs/hep-ph/0105008}{{\tt hep-ph/0105008}}].

\bibitem{quarkonium1}
N.~Brambilla, Y.~Sumino, and A.~Vairo, {\it Quarkonium spectroscopy and
  perturbative QCD: A new perspective},  {\em Phys. Lett.} {\bf B513} (2001)
  381--390, [\href{http://xxx.lanl.gov/abs/hep-ph/0101305}{{\tt
  hep-ph/0101305}}].

\bibitem{quarkonium2}
N.~Brambilla, Y.~Sumino, and A.~Vairo, {\it Quarkonium spectroscopy and
  perturbative qcd: Massive quark loop effects},  {\em Phys. Rev.} {\bf D65}
  (2002) 034001, [\href{http://xxx.lanl.gov/abs/hep-ph/0108084}{{\tt
  hep-ph/0108084}}].

\bibitem{quarkonium3}
N.~Brambilla {\em et~al.}, {\it Heavy quarkonium physics},
  \href{http://xxx.lanl.gov/abs/hep-ph/0412158}{{\tt hep-ph/0412158}}.


\bibitem{hoangfull}
A.~H. Hoang, {\it Bottom quark mass from upsilon mesons: Charm mass effects},
  \href{http://xxx.lanl.gov/abs/hep-ph/0008102}{{\tt hep-ph/0008102}}.

\bibitem{babarhadmom}
{\bf BABAR} Collaboration, B.~Aubert {\em et~al.}, {\it Measurements of moments
  of the hadronic mass distribution in semileptonic B decays},  {\em Phys.
  Rev.} {\bf D69} (2004) 111103,
  [\href{http://xxx.lanl.gov/abs/hep-ex/0403031}{{\tt hep-ex/0403031}}].

\bibitem{bellehadmom}
{\bf BELLE} Collaboration, K.~Abe {\em et~al.}, {\it Moments of the hadronic
  mass spectrum in inclusive semileptonic B decays at belle},
  \href{http://xxx.lanl.gov/abs/hep-ex/0408139}{{\tt hep-ex/0408139}}.

\bibitem{neubertvub}
F.~De~Fazio and M.~Neubert, {\it $ B \to X_u l \bar{\nu}$
 decay distributions to
  order $\alpha_s$},  {\em JHEP} {\bf 06} (1999) 017,
  [\href{http://xxx.lanl.gov/abs/hep-ph/9905351}{{\tt hep-ph/9905351}}].

\bibitem{ossola}
P.~Gambino, G.~Ossola, and N.~Uraltsev, {\it Hadronic mass and $q^2$ 
moments of
  charmless semileptonic B decay distributions},  {\em JHEP} {\bf 09} (2005)
  010, [\href{http://xxx.lanl.gov/abs/hep-ph/0505091}{{\tt hep-ph/0505091}}].

\bibitem{charmlesstheory}
B.~O. Lange, M.~Neubert, and G.~Paz, {\it Theory of charmless inclusive B
  decays and the extraction of $V_{ub}$},
  \href{http://xxx.lanl.gov/abs/hep-ph/0504071}{{\tt hep-ph/0504071}}.

\bibitem{charmlessbabar}
{\bf BABAR} Collaboration, B.~Aubert {\em et~al.}, {\it Measurement of the
  partial branching fraction for inclusive charmless semileptonic B decays and
  extraction of $V_{ub}$},  \href{http://xxx.lanl.gov/abs/hep-ex/0507017}{{\tt
  hep-ex/0507017}}.

\bibitem{vubbabarpaper}
{\bf BABAR} Collaboration, B.~, Aubert {\em et~al.}, {\it Measurement of the
  inclusive electron spectrum in charmless semileptonic B decays near the
  kinematic endpoint and determination of $V_{ub}$|},
  \href{http://xxx.lanl.gov/abs/hep-ex/0509040}{{\tt hep-ex/0509040}}.

\bibitem{gardicharmless}
J.~R. Andersen and E.~Gardi, {\it Inclusive spectra in charmless semileptonic B
  decays by dressed gluon exponentiation},
  \href{http://xxx.lanl.gov/abs/hep-ph/0509360}{{\tt hep-ph/0509360}}.

\bibitem{becher}
T.~Becher and R.~J. Hill, {\it Comment on form factor shape and extraction of
  $V_{ub}$ from $B \to \pi l \nu$},
  \href{http://xxx.lanl.gov/abs/hep-ph/0509090}{{\tt hep-ph/0509090}}.

\bibitem{lukephoton}
A.~F. Falk, M.~E. Luke, and M.~J. Savage, {\it Nonperturbative contributions to
  the inclusive rare decays $B \to X_s \gamma$ and $B \to X_s l^+
  l^-$},  {\em Phys. Rev.} {\bf D49} (1994) 3367--3378,
  [\href{http://xxx.lanl.gov/abs/hep-ph/9308288}{{\tt hep-ph/9308288}}].

\bibitem{anatomy}
A.~L. Kagan and M.~Neubert, {\it QCD anatomy of $B \to X_s \gamma$ decays},
  {\em Eur. Phys. J.} {\bf C7} (1999) 5--27,
  [\href{http://xxx.lanl.gov/abs/hep-ph/9805303}{{\tt hep-ph/9805303}}].

\bibitem{gardi}
J.~R. Andersen and E.~Gardi, {\it Taming the $B \to X_s \gamma$ spectrum by
  dressed gluon exponentiation},  {\em JHEP} {\bf 06} (2005) 030,
  [\href{http://xxx.lanl.gov/abs/hep-ph/0502159}{{\tt hep-ph/0502159}}].

\bibitem{bxgbabar}
{\bf BaBar} Collaboration, F.~Bucci {\em et~al.}, {\it Results from the babar
  fully inclusive measurement of $B\to X_s\gamma$},
  \href{http://xxx.lanl.gov/abs/hep-ex/0507001}{{\tt hep-ex/0507001}}.

\bibitem{bxgbelle}
{\bf Belle} Collaboration, P.~Koppenburg {\em et~al.}, {\it An inclusive
  measurement of the photon energy spectrum in $B\to X_s\gamma$ decays},  {\em
  Phys. Rev. Lett.} {\bf 93} (2004) 061803,
  [\href{http://xxx.lanl.gov/abs/hep-ex/0403004}{{\tt hep-ex/0403004}}].

\bibitem{neubertfactorization}
S.~W. Bosch, B.~O. Lange, M.~Neubert, and G.~Paz, {\it Factorization and
  shape-function effects in inclusive B meson decays},  {\em Nucl. Phys.} {\bf
  B699} (2004) 335--386, [\href{http://xxx.lanl.gov/abs/hep-ph/0402094}{{\tt
  hep-ph/0402094}}].

\bibitem{shapemanohar}
C.~W. Bauer and A.~V. Manohar, {\it Shape function effects in b --> x/s gamma
  and b --> x/u l nu decays},  {\em Phys. Rev.} {\bf D70} (2004) 034024,
  [\href{http://xxx.lanl.gov/abs/hep-ph/0312109}{{\tt hep-ph/0312109}}].


\bibitem{shape2}
I.~Bizjak, A.~Limosani, and T.~Nozaki, {\it Determination of the  B quark
  leading shape function parameters in the shape function scheme using the
  belle $B\to X_s \gamma$ photon energy spectrum},
  \href{http://xxx.lanl.gov/abs/hep-ex/0506057}{{\tt hep-ex/0506057}}.

\end{thebibliography}
\end{document}